\documentclass[aps,prl,showpacs,twocolumn,floats,epsfig,pdflatex]{revtex4}
\usepackage{amssymb}
\usepackage{amsbsy}
\usepackage{amsmath}
\usepackage{epsfig}
\usepackage{graphicx}
\begin{document}

\title {Surface-to-bulk scattering in topological insulator films}
\author{Kush Saha}
\author{Ion Garate}
\affiliation{D\'epartement de Physique and Regroupement Qu\'eb\'ecois sur les Mat\'eriaux de Pointe, Universit\'e de Sherbrooke, Sherbrooke, Qu\'ebec, Canada J1K 2R1}
\date{\today}
\begin{abstract}
We present a quantitative microscopic theory of the disorder- and phonon-induced coupling between surface and bulk states in topological insulator (TI) films.
We find a simple mathematical structure for the surface-to-bulk scattering matrix elements and confirm the importance of bulk-surface coupling in transport and photoemission experiments, assessing its dependence on temperature, carrier density, film thickness and particle-hole asymmetry.
\end{abstract}
\maketitle

{\em Introduction.--}
The advent of three dimensional topological insulators (TIs)~\cite{ti} has ignited a race to develop novel quantum devices capable of exploiting the peculiar transport and magnetoelectric properties of these materials~\cite{ti2}. 
A central problem in the field of TI devices is to probe and isolate their topological surface states through electrical transport measurements. 
In practice, this is a difficult task because all topological ``insulators'' contain residual bulk carriers that couple to surface states through disorder and phonons.
Because such bulk-surface coupling is undesirable for device applications, it is important to understand it in detail.

The role of bulk-surface coupling in TI films has been amply documented in magnetotransport experiments~\cite{transport}. 
Yet, there is no good microscopic understanding of it.
The objective of this paper is to fill this void by presenting a microscopic theory of disorder and phonon-induced bulk-surface coupling.
What are the parameters that govern it?
How does it depend on temperature, carrier concentration and film thickness?
These questions, addressed by our theory,  have implications not only for electrical transport on TI surfaces, but also for tunneling microscopy~\cite{kim0}, photoemission~\cite{photo}, optical spectroscopy~\cite{optics}, relaxation of hot electrons~\cite{hot} and spintronics~\cite{hugh}.

\begin{figure}
\rotatebox{0}{\includegraphics*[width=\linewidth]{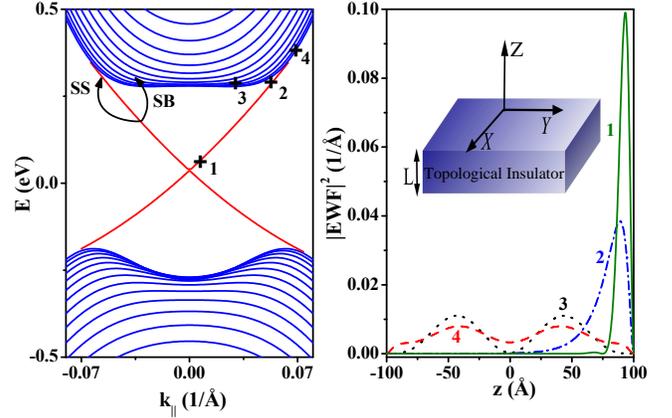}}
\caption{(Color online) {\it Left:} Calculated bulk and surface energy bands for a pristine Bi$_2$Se$_3$ film of thickness $L=20\, {\rm nm}$, using the band parameters given in Ref.~[\onlinecite{zhang}]. 
Each band is doubly degenerate.
For non-ultrathin films, the two degenerate surface states can be chosen to be localized on opposite surfaces.
The surface and bulk states coexist in an energy interval near the bulk band edges.
The zero of energy corresponds to the middle of the bulk bandgap at $k_\parallel=0$.
The arrows labelled as SS and SB describe examples of (inelastic) intrasurface and surface-to-bulk scattering.
{ \it Right:} Envelope wave function (EWF) profiles for some bulk and surface states (the numbers 1-4 labelling them coincide with those of (a)). 
The surface state, whose penetration depth grows with energy, is indistinguishable from a bulk state when its penetration depth becomes comparable to $L$. 
{\em Inset:} scheme of a TI film. 
} 
\label{fig:fig1}
\end{figure}

{\em Model.--}
We consider a TI film of thickness $L$,  whose surfaces are perpendicular to the $z$ direction (Fig.~\ref{fig:fig1}).
For a pristine film of non-interacting electrons, the low-energy electronic structure can be obtained from the Hamiltonian~\cite{zhang,liu,zhou,shan}  
\begin{align}
\label{eq:hm0}
{\cal H}({\bf k}_\parallel,\partial_z)=\epsilon_{{\bf k}_\parallel,\partial_z}+M_{{\bf k}_\parallel,\partial_z} \tau^z +{\bf d}_{{\bf k}_\parallel,\partial_z}\cdot{\boldsymbol\sigma}\tau^x,
\end{align}
where $\sigma^i$ and $\tau^i$ are Pauli matrices in spin and orbital space (respectively), ${\bf k}_\parallel=(k_x,k_y)$ is the momentum parallel to the surfaces, 
$\epsilon_{{\bf k}_\parallel,\partial_z}=\gamma_\parallel k_\parallel^2-\gamma_{\perp}\partial_{z}^2 $ is the particle-hole symmetry-breaking term, $M_{{\bf k}_\parallel,\partial_z}=m-\beta_{\parallel} k_\parallel^2+\beta_{\perp} \partial_z^2$ is the Dirac/Schr\"odinger mass term, $d^x_{{\bf k}_\parallel,\partial_z}=v_\parallel k_x$, $d^y_{{\bf k}_\parallel,\partial_z}=v_\parallel k_y$ and $d^z_{{\bf k}_\parallel,\partial_z}=-i v_\perp \partial_z$.
The band parameters $v_{\parallel, \perp}$, $\gamma_{\parallel,\perp}$, $\beta_{\parallel, \perp}$ and $m$ are material-dependent (although $m/\beta_\perp>0$ is required for a TI).

The eigenstates of Eq.~(\ref{eq:hm0}) can be written as
\begin{equation}
\label{eq:Psi}
\phi_{{\bf k}_\parallel \alpha}({\bf r})=e^{i {\bf k}_\parallel\cdot{\bf r}_\parallel} u_{{\bf k}_\parallel \alpha}({\bf r})/\sqrt{A},
\end{equation}
where $A$ is the sample area in the $xy$ plane, $\alpha$ labels surface and bulk bands (hereafter referred to as $\alpha\in S$ and $\alpha\in B$, respectively)
and $u_{{\bf k}_\parallel \alpha}({\bf r})$ has the periodicity of the lattice in the $xy$ plane.
All eigenstates obey the boundary conditions $\phi_{{\bf k}_\parallel \alpha}({\bf r}_\parallel,z=\pm L/2)=0$ and the orthogonality condition $\langle\phi_{{\bf k}_\parallel \alpha}|\phi_{{\bf k}'_\parallel \alpha'}\rangle=\delta_{\alpha \alpha'}\delta_{{\bf k}_\parallel {\bf k}'_\parallel}$.
The eigenenergies  are denoted as $E_{{\bf k}_\parallel \alpha}$ and displayed in Fig.~\ref{fig:fig1}.

\begin{figure}
\includegraphics[height=4.2cm, width=7.4cm]{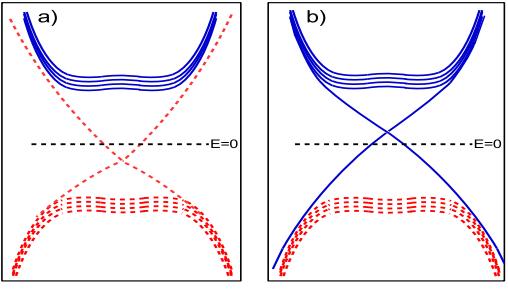} 
\caption{(Color online) Cartoon representing the average value of the orbital pseudospin $\tau^z$ in low-energy energy bands.
Dashed (red) and solid (blue) lines indicate $\langle\tau^z\rangle<0$ and $\langle\tau^z\rangle>0$, respectively.
At higher momenta (not shown), the bulk $\langle\tau^z\rangle$ reverses sign (this is a direct consequence of band inversion), but the surface $\langle\tau^z\rangle$ does not.
Panels (a) and (b) have opposite signs of $\gamma_\perp$.
The SB scattering rate is enhanced if the bulk and surface states near the Fermi level have parallel pseudospin orientation (this happens for $p-$doped samples in (a) and $n-$doped samples in (b)).} 
\label{fig:el2}
\end{figure}

{\em Surface-to-bulk (SB) scattering rate.--}
Consider an electron in surface band $\alpha$, with momentum ${\bf k}_\parallel$ and energy $E_{{\bf k}_\parallel \alpha}=\epsilon_F$, where $\epsilon_F$ is the Fermi energy.
In presence of dilute impurities and weak electron-phonon interactions, this surface electron scatters into the bulk at a rate $\Gamma_{{\bf k}_\parallel \alpha, SB}=\Gamma^{\rm imp}_{{\bf k}_\parallel \alpha, SB}+\Gamma^{\rm ph}_{{\bf k}_\parallel \alpha, SB}$, where~\cite{sm}
\begin{align}
\label{eq:rate}
\Gamma_{{\bf k}_\parallel \alpha,{\rm SB}}^{\rm imp} (\epsilon_F) &=2\pi\sum_{{\bf k}'}\sum_{\alpha'\in{\rm B}} |F^{\rm imp}_{{\bf k}_\parallel \alpha, {\bf k}' \alpha'}|^2 \delta(\xi_{{\bf k}'_\parallel \alpha'})\nonumber\\
\Gamma_{{\bf k}_\parallel \alpha,{\rm SB}}^{\rm ph} (\epsilon_F) &=2\pi\sum_{{\bf k}'}\sum_{\alpha'\in{\rm B}}\sum_{s=\pm} |F^{\rm ph}_{{\bf k}_\parallel \alpha, {\bf k}' \alpha'}|^2 \left[n(s\,\xi_{{\bf k}'_\parallel \alpha'})\right.\nonumber\\
&\left.+f(s\,\xi_{{\bf k}'_\parallel \alpha'})\right]\delta(s\,\xi_{{\bf k}'_\parallel \alpha'} - \omega_{{\bf k}_\parallel-{\bf k}'_\parallel, k'_z})
\end{align}
are the disorder- and phonon-induced SB scattering rates, ${\bf k}'=({\bf k}'_\parallel, k'_z)$, $s=\pm 1$ labels phonon absorption and emission processes, $\omega_{\bf k}$ is the phonon frequency, $\xi_{{\bf k} \alpha}=E_{{\bf k} \alpha}-\epsilon_F$, $n(\epsilon)$ is the phonon occupation factor, $f(\epsilon)$ is the electron occupation factor, $\delta$ is the Dirac delta ensuring energy conservation,  and
\begin{equation}
\label{eq:me}
F^{\rm imp (ph)}_{{\bf k}_\parallel \alpha, {\bf k}' \alpha'}\equiv g^{\rm imp (ph)}_{{\bf k}_\parallel-{\bf k}'_\parallel, k'_z} \langle u_{{\bf k}_\parallel \alpha}|e^{i k'_z z}|u_{{\bf k}'_\parallel \alpha'}\rangle/A
\end{equation}
is the SB scattering matrix element that reflects the wave function overlap between bulk and surface states.
Here, $g_{\bf k}^{\rm imp}$ ($g_{\bf k}^{\rm ph}$) is the electron-impurity (electron-phonon) coupling.
For elastic point scatterers, $|g_{\bf k}^{\rm imp}|^2$ is independent of ${\bf k}$ and proportional to the impurity concentration~\cite{mahan}.
For acoustic phonons,  $|g_{\bf k}^{\rm ph}|^2=C_{\rm ac}^2 k^2/(2\rho V \omega_{\bf k})$, where $C_{\rm ac}$ is the acoustic deformation potential and $\rho$ is the atomic density.
For optical phonons, $|g_{\bf k}^{\rm ph}|^2=C_{\rm op}^2/(2\rho V \omega_0)$, where $C_{\rm op}$ is the optical deformation potential and $\omega_0$ is the optical phonon frequency.

The surface-to-surface (SS) scattering rate, denoted as $\Gamma_{{\bf k}_\parallel n, SS}$ and used below for comparative purposes, can be calculated from Eq.~(\ref{eq:rate}) by summing $\alpha'$ over surface (rather than bulk) bands.
For the TI films considered in this work ($L\gtrsim 10\, {\rm nm}$), the scattering between {\em opposite} surfaces of the film is negligible and thus $\Gamma_{{\bf k}_\parallel \alpha, SS}$ describes the intrasurface scattering rate.

The derivation of Eq.~(\ref{eq:rate}) relies on the envelope function approximation and thus implicitly assumes that the surface potential, the impurity potential and the electron-phonon interaction are smooth at the atomic lengthscale.
Also, for concreteness we consider the scattering of surface electrons off non-magnetic impurities and bulk~\cite{stroscio} phonons.
In thin films grown on electrically insulating substrates,  the neglect of phonon quantization effects is not incompatible with the incorporation of electronic quantization effects.  
Moreover, the key ideas of this work are independent of the details of phonons.

Roughly speaking, $\Gamma_{\rm SB}$ is determined by the convolution between the bulk density of states and the square of the SB scattering matrix elements.
The former implies that the SB scattering rate will be enhanced at the van Hove singularities of the bulk density of states.
At first glance, the matrix elements are complicated and it is unclear whether there are simple principles governing them.
At a closer look~\cite{sm}, we find that it is possible to gain valuable physical intuition by rewriting Eq.~(\ref{eq:me}) as
\begin{align}
\label{eq:tt}
|F^{\rm imp (ph)}_{{\bf k}_\parallel \alpha, {\bf k}' \alpha'}|^2 &\propto 1 +\zeta_{k_z' \alpha'}\langle\tau^z\rangle^S_{{\bf k}_\parallel \alpha} \langle\tau^z\rangle^B_{{\bf k}'_\parallel \alpha'}\nonumber\\
&+\zeta'_{k_z' \alpha'}\langle{\boldsymbol\sigma}\tau^x\rangle^S_{{\bf k}_\parallel \alpha}\cdot\langle{\boldsymbol\sigma}\tau^x\rangle^B_{{\bf k}'_\parallel \alpha'},
\end{align}
where $\langle O\rangle^{S (B)}_{{\bf k} n} \equiv\langle u_{{\bf k} \alpha}|O| u_{{\bf k} \alpha}\rangle/A$ denotes the expectation value of an operator $O$ for an electron with momentum ${\bf k}$ in band $\alpha\in S$ ($\alpha\in B$).
The real numbers $\zeta_{k_z' \alpha'}$ and $\zeta'_{k_z' \alpha'}$ depend on the overlap of the bulk and surface wave functions~\cite{sm} but are independent of ${\bf k}_\parallel$, ${\bf k}'_\parallel$, the surface band index and of whether $\alpha'$ is in the conduction or valence band.
 
The main idea from Eq.~(\ref{eq:tt}) is that the SB coupling is sensitive to the relative orientation of the surface and bulk expectation values for the orbital pseudospin $\tau^z$ and the entangled spin-orbital operator $\sigma^i\tau^x$.
The reason why only these operators appear in Eq.~(\ref{eq:tt}) is that there are no other operators for which the surface and bulk expectation values are {\em simultaneously} nonzero (as evidenced by the matrix structure of Eq.~(\ref{eq:hm0})).
For example, the SB coupling is insensitive to the spin polarization of the surface states because the bulk states are spin-unpolarized. 

Let us concentrate on the second term in the rhs of Eq.~(\ref{eq:tt}). 
Assuming non-ultrathin films and $|\gamma_\perp|\leq |\beta_\perp|$, we find
\begin{equation}
\langle\tau^z\rangle_{{\bf k}_\parallel \alpha}^S\simeq\gamma_\perp/\beta_\perp\,\,\,;\,\,\, \langle\tau^z\rangle_{{\bf k}'_\parallel \alpha'}^B \simeq M_{{\bf k}'_\parallel, \lambda}/E_{{\bf k}'_\parallel \alpha'},
\end{equation}
where $\lambda$ is the quantized momentum along the $z$-direction in bulk band $\alpha'$.
Remarkably, the particle-hole asymmetry parameter $\gamma_{\perp}$ ``polarizes'' the surface orbitals along $z$. 
This polarization is proportional to $m \gamma_\perp/\beta_\perp$, which is the shift of the Dirac point away from the middle of the bulk bandgap.
In addition, $\langle\tau^z\rangle^S_{{\bf k}_\parallel \alpha}$ is largely independent of ${\bf k}_{\parallel}$ and $\alpha$~\cite{sm}, i.e. it is the same for the upper and lower Dirac cones.
In stark contrast, $\gamma_\perp$ plays no role in the pseudospin polarization of the bulk bands (because $\gamma_\perp$ multiplies an identity matrix in Eq.~(\ref{eq:hm0})) and $\langle\tau^z\rangle^B$ changes sign from the conduction to the valence band.
Consequently, when $|\gamma_\perp/\beta_\perp|\sim O(1)$ (i.e. when the Dirac point is close to a bulk band edge), Eq.~(\ref{eq:tt}) predicts a strong asymmetry on the magnitude of SB scattering matrix element between hole-doped and electron-doped films (cf. Fig.~\ref{fig:el2}).
In Bi$_2$Se$_3$, this asymmetry is more pronounced when the surface under consideration is perpendicular to the quintuple layers, because $|\gamma_\perp/\beta_\perp| < |\gamma_\parallel/\beta_\parallel|$.

The third term in the rhs of Eq.~(\ref{eq:tt}) can be interpreted similarly. 
However, this term is less important than the one preceding it because its contribution to the SB scattering rate vanishes by symmetry in the cases of point disorder and optical phonons.

\begin{figure}
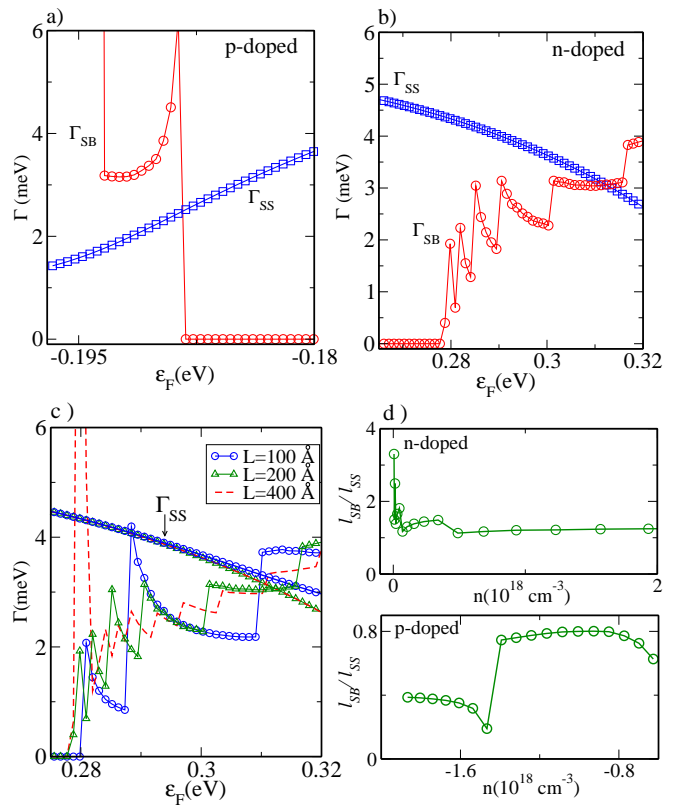

\rotatebox{0}{\includegraphics*[width=\linewidth]{fig3.eps}}\\
\rotatebox{0}{\includegraphics*[width=\linewidth]{fig3a.eps}}
\caption{
(Color online) (a)  Elastic (zero-temperature) SB scattering rate in a $20\,{\rm nm}$-thick hole-doped Bi$_2$Se$_3$ film, calculated numerically from Eq.~(\ref{eq:rate}). 
The disorder strength is chosen such that the surface state mean free path is about $50 {\rm nm}$.
$\Gamma_{\rm SB}$ vanishes unless $\epsilon_F$ intersects bulk bands.
For comparison, the SS scattering rate is also shown. 
$\Gamma_{\rm SS}$ decreases with $|\epsilon_F|$ because the increase in the density of states is overcompensated by a reduction in the SS scattering matrix element. 
(b) Same as (a), but for an electron-doped film. 
(c) Same as (b), but for different film thicknesses.
(d) (Top) Ratio between the elastic SB and SS scattering lengths for a  $20\,{\rm nm}$-thick Bi$_2$Se$_3$ film as function of electron density.
The SB (SS) scattering length is defined via $l_{\rm SB (SS)}=\sqrt{D \tau_{\rm SB (SS)}}$, where $D$ is the diffusion constant and $\tau_{\rm SB (SS)}=1/\Gamma_{\rm SB (SS)}$.
{(Bottom)} Same plot, as a function of hole density.
For a typical value of $l_{\rm SS}$, $l_{\rm SB}$ can be comparable to the typical phase relaxation length $l_\phi$ in weakly doped samples.
For more highly doped films, $l_{\rm SB}\ll l_\phi$.}
\label{fig:el1}
\end{figure}

{\em Elastic bulk-surface coupling.--}
At low temperatures, phonon-induced SB scattering is suppressed and $\Gamma_{\rm SB}\simeq \Gamma^{\rm imp}_{SB}$.
Using Eq.~(\ref{eq:tt}), the first line of Eq.~(\ref{eq:rate}) turns into~\cite{sm}
\begin{equation}
\label{eq:simp}
\Gamma_{{\bf k}_\parallel \alpha,{\rm SB}} (\epsilon_F)\simeq |g^{\rm imp}_{\rm eff}|^2 \nu_B(\epsilon_F) \left(1+\langle\tau^z\rangle^S_{{\bf k}_\parallel \alpha}\langle\langle\tau^z\rangle\rangle^B_{\epsilon_F}\right),
\end{equation}
where $g^{\rm imp}_{\rm eff}\propto g^{\rm imp}$ is an effective electron-disorder coupling, $\nu_B(\epsilon_F)$ is the bulk density of states at the Fermi level (per unit volume) and
\begin{equation}
\label{eq:ave}
\langle\langle\tau^z\rangle\rangle^B_{\epsilon_F} \equiv\frac{1}{\nu_B (\epsilon_F)}\frac{1}{A L}\sum_{{\bf k}'_\parallel \alpha'} \zeta_{\alpha'}\langle\tau^z\rangle^B_{{\bf k}'_\parallel \alpha'}\delta(\xi_{{\bf k}'_\parallel \alpha'})
\end{equation}
is the Fermi surface average of the $z-$component of the bulk orbital pseudospin (weighted with a factor $\zeta_{\alpha'}$).
Hence, the SB scattering rate is (i) sensitive to the relative pseudospin orientation of the surface and bulk states at the Fermi level, and (ii) proportional to the bulk density of states at the Fermi level.
The factor $\zeta_{\alpha'}$ is positive, independent of $\gamma_{\perp,\parallel}$ and independent of whether $\alpha'$ is a conduction or valence band~\cite{sm}. 
Near the bulk band edges, $\zeta_{\alpha'}$ depends weakly on $\alpha'$.
Because $\zeta_{\alpha'}>0$, the SB scattering rate is greater if the bulk and orbital pseudospins are {\em aligned}.
As mentioned above, this effect is especially prononunced if the surface Dirac point is close to a bulk band edge.

Figure~\ref{fig:el1} illustrates the elastic SB scattering rate for Bi$_2$Se$_3$. 
On one hand, there is a sequence of van Hove singularities as a function of the Fermi level, because the energy dispersions of quantum well states near the bulk band edges have extrema at $k_\parallel\neq 0$ (cf. Fig.~\ref{fig:fig1}).
On the other hand, the SB scattering rate grows as the Fermi level delves deeper into the bulk conduction or valence bands. 
This is due to two reasons. First, the number of bulk states available for scattering is greater for larger $|\epsilon_F|$.
Second, the wave function overlap between bulk and surface states is stronger at higher energy because the penetration depth of the surface states increases therein~\cite{sm}.

For comparison, Fig.~\ref{fig:el1} also shows the surface-to-surface (SS) scattering rate.
Since the wave function overlap between a surface state and a bulk state is weaker than the overlap of a surface state with itself, the SB scattering rate is often lower than the SS scattering rate.
However, this difference decreases gradually with increased bulk doping and turns insignificant when the penetration depth of the ``surface state'' at the Fermi level becomes comparable to the film thickness.
Furthermore, the van Hove singularities in the bulk density of states can overcompensate for the deficit in the SB matrix elements, especially if the Fermi level is in the valence band.
In Fig.~\ref{fig:el1}c, we show that the SS scattering rate is largely independent of film thickness (for $L\gtrsim 10 {\rm nm}$) and that the SB scattering rate depends on $L$ through a rearrangement in the sequence of the van Hove singularities.

Our results have implications for low-field magnetoresistance experiments in TI films~\cite{transport,bardarson}.
These experiments are often interpreted under the belief that the SB scattering rate far exceeds the phase relaxation rate in doped TIs without depletion layers.
Although Fig.~\ref{fig:el1} corroborates this assumption for Bi$_2$Se$_3$ in all but very weakly doped samples, it also adds two generic insights that have gone unnoticed thus far.
First, the bulk-surface coupling depends sensitively on the carrier density and on the film thickness because the van Hove singularities shift as a function of these.
In order to discern these singularities experimentally, the electronic mean free path must exceed the film thickness.
Second, the bulk-surface coupling also depends on the degree to which particle-hole symmetry is broken: the coupling is stronger in the bulk band that is closest to the surface Dirac point.

\begin{figure}
\rotatebox{0}{\includegraphics*[width=\linewidth]{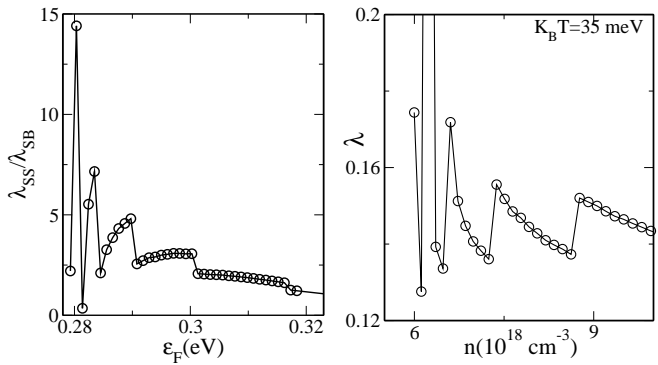}}
\caption{{\em Left:} High-temperature ratio between the inelastic SS and SB contributions to the Fermi-level electron-phonon coupling constant $\lambda$, calculated numerically from Eq.~(\ref{eq:rate}) for a $20\,{\rm nm}$-thick Bi$_2$Se$_3$ film.
The phonon frequency is $\omega_0=30 {\rm meV}$. 
{\em Right:} $\lambda$ as a function of carrier density.}
 \label{fig:inel1}
\end{figure}

{\em Inelastic bulk-surface coupling.--}
At high temperature, the SB scattering is dominated by phonons and hence $\Gamma_{\rm SB}\simeq \Gamma^{\rm ph}_{\rm SB}$.
The phonon-induced SB scattering rate is nonzero even in the insulating bulk regime,  provided that $\epsilon_F$ lies within $\omega_0$ from the bulk band edge (for optical phonons) or within a Debye frequency from the bulk band edge (for acoustic phonons).
For simplicity we concentrate on optical phonons, which are known to play an important role in TIs~\cite{becker}.
Then, Eq.~(\ref{eq:rate}) yields~\cite{sm}
\begin{align}
\label{eq:simph}
&\Gamma_{{\bf k}_\parallel \alpha,{\rm SB}} (\epsilon_F)\simeq |g^{\rm ph}_{\rm eff}|^2 \nu_B(\epsilon_{F+})\left[n(\omega_0)+f(\omega_0)\right]\nonumber\\
&\times\left(1+\langle\tau^z\rangle^S_{{\bf k}_\parallel \alpha}\langle\langle\tau^z\rangle\rangle^B_{\epsilon_{F+}}\right) +(\epsilon_{F+}\to\epsilon_{F-}),
\end{align}
where $\epsilon_{F\pm}=\epsilon_F\pm\omega_0$, $g^{\rm ph}_{\rm eff}\propto g^{\rm ph}$ is an effective electron-phonon coupling and $\langle\langle\tau^z\rangle\rangle^B$ was defined in Eq.~(\ref{eq:ave}).
The contribution from acoustic phonons has the same form, provided that the quasi-elastic approximation holds~\cite{sm}.
Comparing Eqs.~(\ref{eq:simp}) and (\ref{eq:simph}), it is apparent that the main ideas from the preceding section hold here as well.

At temperature $T\gtrsim\omega_0$, the measurement of the Fermi-level surface-state broadening $\Gamma$ in photoemission experiments~\cite{hatch,pan,zhu, crepaldi,li} allows to determine the dimensionless electron-phonon coupling constant $\lambda$, defined as $\lambda=\Gamma/(2\pi k_B T)$~\cite{chulkov}.
The knowledge of this quantity is crucial to interpret transport and optical properties, as well as possible superconducting instabilities. 
For doped Bi$_2$Se$_3$, one group claims an ``exceptionally'' small value of  $\lambda\simeq 0.1$, while others disagree.
Bulk-surface coupling could be a relevant actor in this controversy.
Since $\Gamma=\Gamma_{\rm SS}+\Gamma_{\rm SB}$, we may write $\lambda=\lambda_{\rm SS} (1+\Gamma_{\rm SB}/\Gamma_{\rm SS})$, where $\lambda_{\rm SS}$ is the intrinsic surface contribution~\cite{giraud} to $\lambda$.
In thin films, due to van Hove singularities in the SB scattering rate, $\lambda$ is much more sensitive than $\lambda_{\rm SS}$ to the carrier density (cf. Fig.~\ref{fig:inel1}) and to the film thickness.
Consequently, the measured $\lambda$ varies significantly even between samples of similar characteristics (cf. Fig.~\ref{fig:inel1}), and can reach $\simeq 0.2$ even when $\lambda_{\rm SS}\simeq 0.1$.

At temperature $T<\omega_0$, optical phonons are exponentially suppressed and $\Gamma^{\rm ph}_{\rm SB}$  is dominated by acoustic phonons.
If $\epsilon_F$ is within the bulk gap and a distance $\Delta$ away from the bulk band edge, $\Gamma^{\rm ph}_{\rm SB}/\Gamma^{\rm ph}_{\rm SS}\propto \exp(-\Delta/T)$ at $T<\Delta$.
In contrast, if $\epsilon_F$ intersects one or more bulk bands, we find~\cite{sm} $\Gamma^{\rm ph}_{\rm SB}\propto T^3$ for $c_s q_-\ll T\ll 2 c_s k_F$. 
Here, $c_s$ is the sound velocity, $q_-$ is the distance (in momentum space) between the surface band and the nearest bulk band at the Fermi level, and $k_F$ is the Fermi momentum on the surface band.    
In this range of temperature, $\Gamma^{\rm ph}_{\rm SB}/\Gamma^{\rm ph}_{\rm SS}$ is independent of $T$.
When $T<c_s q_-$, $\Gamma^{\rm ph}_{\rm SB}$ vanishes while $\Gamma^{\rm ph}_{\rm SS}$ still varies as $T^3$.
At low enough temperature, the inelastic SB scattering is dominated by Coulomb interactions, not included herein.  

{\em Conclusions.--}
We have presented the first microscopic theory of bulk-surface coupling in topological insulator films and have distilled simple guiding principles that govern it.
Two of our predictions may help interpret transport and photoemission experiments in Bi$_2$Se$_3$ and related materials:
(i) the bulk-surface coupling is sensitive to the relative orientation of the orbital pseudospin of the bulk and surface states,  
(ii) the bulk-surface coupling can alter the measured electron-phonon coupling of the surface states at the Fermi level by up to 50 \%.

{\em Acknowledgements.--}
This work has been funded by U. de Sherbrooke, Qu\'ebec's RQMP and Canada's NSERC.
The numerical calculations were performed on computers provided by Calcul Qu\'ebec and Compute Canada.

\begin{widetext}
\section{Supplementary material}

\subsection{A.~Phonon-induced scattering rate in a topological insulator film}

The objective of this section is to derive the second and third lines of Eq.~(3) in the main text.
We begin with the simplest Hamiltonian for the electron-phonon interaction:
\begin{equation}
\label{ep1}
H_{ep}=\int d{\bf r} \rho({\bf r}) \Phi({\bf r}), 
\end{equation}  
where $\rho({\bf r})=\Psi({\bf r})^{\dagger}\Psi({\bf r})$ is the electron density operator, 
\begin{align}
\Phi({\bf r})=\sum_{\bf q} e^{i {\bf q}\cdot\bf{r}} g^{\rm ph}_{\bf q}  {\bf e}_{\bf q} (a_{\bf q}+a_{-\bf q}^{\dagger})
\end{align}
is the potential generated by long wavelength lattice vibrations, $g^{\rm ph}_{\bf q}$ is the electron-phonon coupling (with units of energy), $a_{\bf q}$ is an operator that annihilates a phonon with momentum ${\bf q}$  and ${\bf e}_{\bf q}$ is the polarization vector of the vibration mode with momentum ${\bf q}$.
Here, we consider only the coupling between electrons and bulk longitudinal phonons.
The assumption of bulk phonons means that we ignore phonon quantization effects.
The fact that we consider electronic confinement while neglecting phonon confinement is not a contradiction when the topological insulator is grown on a substrate that is electrically insulating but thermally conducting (which is the case in many experiments).

Next, we write $\Psi({\bf r})$ in the eigenstate basis,
\begin{equation}
\label{wav}
\Psi({\bf r})=\sum_{{\bf k_{\parallel}}\alpha}\phi_{{\bf k_{\parallel}}\alpha}({\bf r})c_{{\bf k_{\parallel}}\alpha},
\end{equation}
where  $c_{{\bf k}_\parallel \alpha}$ annihilates an electron in band $\alpha$ (which may be either a surface band or a bulk band) with momentum ${\bf k}_{\parallel}$. 
Then, the density operator can be written as
\begin{eqnarray}
\label{rho}
\rho({\bf r})=\sum_{{\bf k_{\parallel}}\alpha,{\bf k'_{\parallel}}\alpha'}\phi^{\ast}_{{\bf k_{\parallel}}\alpha}({\bf r})\phi_{{\bf k'_{\parallel}}\alpha'}({\bf r})c_{{\bf k_{\parallel}}\alpha}^{\dagger}c_{{\bf k'_{\parallel}}\alpha'}.
\end{eqnarray}
For system with broken translational symmetry in the $z$ direction, $\phi_{{\bf k_{\parallel}}n}({\bf r})$ reads 
\begin{eqnarray}
\label{phi}
\phi_{{\bf k_{\parallel}}\alpha}({\bf r})=\phi_{{\bf k_{\parallel}}\alpha}({\bf r_{\parallel}},z)=\frac{1}{\sqrt{A}}e^{i {\bf k_{\parallel}}.{\bf r_{\parallel}}} u_{{\bf k_{\parallel}}\alpha}({\bf r_{\parallel}},z),
\end{eqnarray}
where $A$ is the area of the crystal in the $xy$ plane. 
Substituting Eq.~(\ref{phi} )in Eq.~(\ref{rho}), we obtain 
\begin{align}
\label{rho1}
\rho({\bf r})=\frac{1}{A}\sum_{{\bf k_{\parallel}}\alpha,{\bf k'_{\parallel}}\alpha'} e^{-i ({\bf k_{\parallel}-k'_{\parallel}}).{\bf r_{\parallel}}} u_{{\bf k_{\parallel}}\alpha}^{*} ({\bf r}) u_{{\bf k'_{\parallel}}\alpha'} ({\bf r})c_{{\bf k_{\parallel}}\alpha}^{\dagger}c_{{\bf k'_{\parallel}}\alpha'}.
\end{align}
In momentum space,
\begin{equation} 
\rho({\bf q})=\int d{\bf r}\, e^{i{\bf q}\cdot{\bf r}} \rho({\bf r})=\frac{1}{A}\sum_{{\bf k_{\parallel}}\alpha,{\bf k'_{\parallel}}\alpha'} \langle u_{{\bf k_{\parallel}}\alpha}|e^{-i ({\bf k_{\parallel}-k'_{\parallel}-q_{\parallel}}).{\bf r_{\parallel}}}e^{i q_z z} |u_{{\bf k'_{\parallel}}\alpha'}\rangle c_{{\bf k_{\parallel}}\alpha}^{\dagger}c_{{\bf k'_{\parallel}}\alpha'}.\label{rhoq}
\end{equation}

Assuming that the surface of the film is not abrupt at the atomic lengthscale, we may invoke the envelope function approximation.
In this approximation, we may expand $u_{{\bf k_{\parallel}}\alpha}({\bf r})$ as
\begin{align}
u_{{\bf k_{\parallel}}\alpha} ({\bf r})=\sum_{j=1}^4 a_{\alpha j}({\bf k}_\parallel, z) u_{{\bf 0} j} ({\bf r}),
\label{unk}
\end{align}
where the coefficients $\{a_{\alpha j}({\bf k}_\parallel, z)\}$ are obtained by diagonalizing Eq.~(1) in the main text and $\{u_{{\bf 0} j} ({\bf r})\}$ are the wave functions corresponding to the basis states.
Here, $j\in\{|P_1\uparrow\rangle,|P_1\downarrow\rangle,|P_2\uparrow\rangle,|P_2\downarrow\rangle\}$, where $\uparrow (\downarrow)$ is the direction of spin and $P_1 (P_2)$ are two orbitals of opposite parity under spatial inversion. 
The functions $a_{\alpha j}({\bf k}_\parallel, z)$ change slowly at the atomic lengthscale, while the functions $u_{{\bf 0} j} ({\bf r})$ change rapidly. 
The normalization condition for the envelope function spinors reads
\begin{equation}
\int_{-L/2}^{L/2} dz \sum_j |a_{\alpha j}({\bf k}_\parallel, z)|^2 =1,
\end{equation}
provided that
\begin{equation}
\frac{1}{V_{\rm cell}} \int_{\rm cell} d {\bf r}\, |u_{{\bf 0} j}({\bf r})|^2=1,
\end{equation}
where $V_{\rm cell}$ is the volume of the unit cell.
It is useful to prove this relation explicitly.
We begin from the normalization condition for the full Bloch's function, $\langle \phi_{{\bf k}_\parallel \alpha} |\phi_{{\bf k}_\parallel \alpha}\rangle = 1$. 
Then,
\begin{align}
&1=\int d {\bf r}\, \phi^*_{{\bf k}_\parallel \alpha} ({\bf r})\phi_{{\bf k}_\parallel \alpha} ({\bf r})=\frac{1}{A}\sum_{jj'}\int d{\bf r_{\parallel}} dz\, a_{j \alpha}^*({\bf k}_\parallel, z) a_{j' \alpha}({\bf k}_\parallel,z) u_{{\bf 0} j}^*({\bf r}) u_{{\bf 0} j'}({\bf r})\nonumber\\
&=\frac{1}{A}\sum_{j j'}\sum_{{\bf R}_\parallel}\sum_Z\int_{\rm cell} d{\bf r_{\parallel}} dz~a_{j \alpha}^*({\bf k}_\parallel, z+Z) a_{j' \alpha}({\bf k}_\parallel,z+Z) u_{{\bf 0} j}^*({\bf r}_\parallel+{\bf R}_\parallel, z+Z) u_{{\bf 0} j'}({\bf r}_\parallel+{\bf R}_\parallel, z+Z)\nonumber\\
&= \frac{N_{\rm cell}}{A}\sum_{j j'}\sum_Z\int_{\rm cell} d{\bf r_{\parallel}} dz~a_{j \alpha}^*({\bf k}_\parallel, z+Z) a_{j' \alpha}({\bf k}_\parallel,z+Z) u_{{\bf 0} j}^*({\bf r}_\parallel, z+Z) u_{{\bf 0} j'}({\bf r}_\parallel, z+Z)\nonumber\\   
&\simeq  \frac{N_{\rm cell}}{A}\sum_{j j'}\sum_Z a_{j \alpha}^*({\bf k}_\parallel, Z) a_{j' \alpha}({\bf k}_\parallel,Z) \int_{\rm cell} d{\bf r_{\parallel}} dz u_{{\bf 0} j}^*({\bf r}_\parallel, z) u_{{\bf 0} j'}({\bf r}_\parallel, z)\nonumber\\   
& =  \frac{N_{\rm cell}}{A}\sum_{j j'}\sum_Z a_{j \alpha}^*({\bf k}_\parallel, Z) a_{j' \alpha}({\bf k}_\parallel,Z) V_{\rm cell} \delta_{j j'}\nonumber\\
& \simeq  \frac{N_{\rm cell} V_{\rm cell}}{A L_{\rm cell}^z}\sum_{j}\int_{-L/2}^{L/2} dZ |a_{j \alpha}({\bf k}_\parallel, Z)|^2 = \sum_{j}\int_{-L/2}^{L/2} dZ |a_{j \alpha}({\bf k}_\parallel, Z)|^2.
\end{align}
In this derivation, we have defined ${\bf R}_\parallel$ and $Z$ as the coordinates that label the centers of the unit cells of the crystal in the $xy$ plane and along the $z$ direction, respectively.
Also, $N_{\rm cell}$ is the number of unit cells in the $xy$ plane (for fixed $z$), $A_{\rm cell}$ is the area of the unit cell in the $xy$ plane, and $L_{\rm cell}^z$ is the length of the unit cell along the $z$ direction. 
Moreover, we have used the envelope wavefunction approximation, which leads to two simplifications.
First,  the variation of $a_{n j}(z)$ within a unit cell is negligible, i.e. $a_{\alpha j}({\bf k}_\parallel, z+Z)\simeq a_{\alpha j}({\bf k}_\parallel, Z)$.
Second, the rapidly varying atomic wave functions are not appreciably altered by the presence of the surface, i.e. $u_{{\bf 0} j} ({\bf r}_\parallel, z+Z)\simeq u_{{\bf 0} j} ({\bf r}_\parallel, z)$. 
Keeping these approximations in mind, we can simplify the matrix element appearing in the expression for $\rho({\bf q})$:
\begin{equation}
\langle u_{{\bf k_{\parallel}}\alpha}|e^{-i ({\bf k_{\parallel}-k'_{\parallel}-q_{\parallel}}).{\bf r_{\parallel}}}e^{i q_z z} |u_{{\bf k'_{\parallel}}\alpha'}\rangle \simeq \delta_{{\bf k}'_\parallel,{\bf k}_\parallel-{\bf q}_\parallel}\langle u_{{\bf k_{\parallel}}\alpha}|e^{i q_z z}|u_{{\bf k'_{\parallel}}\alpha'}\rangle.
\label{matap}
\end{equation}
Here, we have used $\sum_{{\bf R}_\parallel} \exp(i {\bf p}_\parallel\cdot{\bf R}_\parallel) = N_{\rm cell} \sum_{{\bf G}_\parallel} \delta_{{\bf p}_\parallel, {\bf G}_\parallel}$, where ${\bf G}_\parallel$ is the reciprocal lattice vector in the $xy$ plane, and we have ignored Umklapp ($G\neq 0$) terms.
The neglect of Umklapp processes is well justified when calculating the surface-to-bulk scattering in weakly doped topological insulators, where the momenta involved are small compared to the size of the Brillouin zone.
 
Combining Eq.~(\ref{rhoq}) with (\ref{matap}), we obtain 
\begin{eqnarray}
\rho({\bf q})=\frac{1}{A}\sum_{{\bf k}_\parallel \alpha,\alpha'} \langle u_{{\bf k_{\parallel}}\alpha}|e^{i q_z z} |u_{{\bf k_{\parallel}-q_{\parallel}}\alpha'}\rangle c_{{\bf k_{\parallel}}\alpha}^{\dagger}c_{{\bf k_{\parallel}-q_{\parallel}}\alpha'}.
\end{eqnarray}
We now treat Eq.~(\ref{ep1}) as a perturbation in the S-matrix expansion of the Green's function (see Ref.~[14] in the main text). 
Keeping only the leading order correction in the expansion, we obtain the self-energy  
\begin{equation}
\label{se}
\Sigma^{\rm ph}_{\alpha {\bf k}_\parallel, \alpha' {\bf k}_\parallel}(i\omega)=-T\sum_{{\bf k'} \alpha'',\nu_m} F^{\rm ph}_{{\bf k}_\parallel \alpha, {\bf k}' \alpha''} (F^{\rm ph})^*_{{\bf k}_\parallel \alpha', {\bf k}' \alpha''}  G^{(0)}_{{\bf k}'_\parallel \alpha''}(i\omega-i\nu_m)D^{(0)}_{{\bf k}_\parallel-{\bf k}'_\parallel, k'_z}(i\nu_m),
\end{equation}
where ${\bf k}'=({\bf k}'_\parallel, k_z')$ (note that we have made a change of variable $q_z\to k_z'$) and 
\begin{equation}
F^{\rm ph}_{{\bf k}_\parallel \alpha, {\bf k}' \alpha'}=g^{\rm ph}_{{\bf k}_\parallel-{\bf k}'_\parallel, k'_z}  \langle u_{{\bf k_{\parallel}}\alpha}|e^{i k'_z z} |u_{{\bf k'_{\parallel}} \alpha'}\rangle/A
\end{equation}
is the electron-phonon scattering matrix element, $G^{(0)}$ and $D^{(0)}$ are the unperturbed electron and phonon Green's functions (respectively), $\omega_l= (2 l+1)\pi T$ ($l\in\mathbb{Z}$) is the fermionic Matsubara frequency at temperature $T$, and  $\nu_m= 2m\pi T$ ($m\in\mathbb{Z}$) is the bosonic Matsubara frequency.
The label $\alpha''$ denotes the intermediate bands that an electron can scatter to and comprises both bulk and surface bands.

In this work we are interested in the transport lifetimes of surface states.
Accordingly, we restrict ourselves to $\alpha=\alpha'\in S$ and we use the abbreviation $\Sigma_{\alpha \alpha}\equiv \Sigma_\alpha$.
In Eq.~(\ref{se}), it is convenient to separate out intermediate states $\alpha''$ into surface and bulk bands,
\begin{equation}
\Sigma^{\rm ph}_{{\bf k}_\parallel \alpha}(i\omega)=\Sigma^{\rm ph}_{{\bf k}_\parallel \alpha, {\rm SS}}(i\omega)+\Sigma^{\rm ph}_{{\bf k}_\parallel \alpha,{\rm SB}}(i\omega).
\end{equation}
After summing over bosonic Matsubara frequencies, we get
\begin{align}
\label{eq:apeqs}
\Sigma^{\rm ph}_{{\bf k}_\parallel \alpha,{\rm SS}}(i\omega)=&\sum_{{\bf k'}\alpha'\in S} |F^{\rm ph}_{{\bf k}_\parallel \alpha, {\bf k}'_\parallel \alpha'}|^2 \left[\frac{1+n(\omega_{{\bf k'_{\parallel}-k_{\parallel}},k'_z})-f (\xi_{{\bf k}'_\parallel \alpha'})}{i\omega-\xi_{{\bf k'_{\parallel}}\alpha'}-\omega_{{\bf k'_{\parallel}-k_{\parallel}},k'_z}}\right.
+\left.\frac{n(\omega_{{\bf k'_{\parallel}-k_{\parallel}},k'_z})+f(\xi_{{\bf k}'_\parallel \alpha'})}{i\omega-\xi_{{\bf k'_{\parallel}}\alpha'} +\omega_{{\bf k'_{\parallel}-k_{\parallel}},k'_z}}\right]\nonumber\\
\Sigma^{\rm ph}_{{\bf k}_\parallel \alpha,{\rm SB}}(i\omega)=&\sum_{{\bf k'}\alpha'\in B} |F^{\rm ph}_{{\bf k}_\parallel \alpha, {\bf k}'_\parallel \alpha'}|^2 \left[\frac{1+n(\omega_{{\bf k'_{\parallel}-k_{\parallel}},k'_z})-f(\xi_{{\bf k}'_\parallel \alpha'})}{i\omega-\xi_{{\bf k'_{\parallel}}\alpha'}-\omega_{{\bf k'_{\parallel}-k_{\parallel}},k'_z}}\right.
+\left.\frac{n(\omega_{{\bf k'_{\parallel}-k_{\parallel}},k'_z})+f(\xi_{{\bf k}'_\parallel \alpha'})}{i\omega-\xi_{{\bf k'_{\parallel}}\alpha'} +\omega_{{\bf k'_{\parallel}-k_{\parallel}},k'_z}}\right],
\end{align}
where $\xi_{{\bf k}\alpha}$ is the electronic energy measured from the Fermi level,  $n(\epsilon)=1/(\exp(\epsilon/T)-1)$ is the phonon occupation factor  and $f(\epsilon)=1/(\exp(\epsilon/T)+1)$ is electron occupation factor. 
Doing the analytical continuation ($i\omega\to\omega+i 0^+$) and taking the imaginary part of the self-energy, we arrive at the phonon-induced scattering rate shown in the main text (Eq.~(3)).
The first line in Eq.~(\ref{eq:apeqs}) gives the intrasurface scattering rate, while the second line gives the surface-to-bulk scattering rate.
In order to obtain scattering rates at the Fermi surface, we set $\omega=0$.

\subsection{B.~Disorder-induced scattering rate in a topological insulator film}

The objective of this section is to derive the first line of Eq.~(3) in the main text.
We consider non-magnetic impurities that are also independent of the orbital quantum number.
This implies that the random impurity potential  $U({\bf r})$ varies slowly at the atomic lengthscale.
The Hamiltonian for electrons interacting with such impurities is
\begin{align}
\label{dis}
H_{\rm imp} &=\int d{\bf r}\, U({\bf r})\rho({\bf r})=
\frac{1}{A}\sum_{{\bf q}} U_{\bf q} \langle u_{{\bf k_{\parallel}}\alpha}|e^{-i({\bf k}_\parallel-{\bf k}'_\parallel-{\bf q}_\parallel)\cdot{\bf r}_\parallel} e^{i q_z z}|u_{{\bf k'_{\parallel}}\alpha'} \rangle\nonumber\\
&= \frac{1}{A}\sum_{{\bf k_{\parallel}} \alpha,{\bf k'} \alpha'}  U_{{\bf k}_\parallel-{\bf k}'_\parallel, k'_z} \langle u_{{\bf k_{\parallel}}\alpha}|e^{i  k'_z z }|u_{{\bf k'_{\parallel}}\alpha'} \rangle  c_{{\bf k_{\parallel}}\alpha}^{\dagger}c_{{\bf k'_{\parallel}}\alpha'},
\end{align}
where we have used the same envelope function approximation of the preceding section.
To second order in ${\cal H}_{\rm dis}$, the electronic Green's function is given by 
\begin{align}
G_{{\bf k}_\parallel \alpha, {\bf k}'_\parallel \alpha'} (i\omega) &=G^{(0)}_{{\bf k}_\parallel \alpha}(i\omega)\, \delta_{\alpha \alpha'} \delta_{{\bf k}_\parallel, {\bf k}'_\parallel}+G^{(0)}_{{\bf k}_\parallel \alpha} (i\omega)\, \delta U_{{\bf k}_\parallel \alpha, {\bf k}'_\parallel \alpha'} \,G^{(0)}_{{\bf k}'_\parallel \alpha'} (i\omega)\nonumber\\ 
&+ G^{(0)}_{{\bf k}_\parallel \alpha} (i\omega)  G^{(0)}_{{\bf k}''_\parallel \alpha''}(i\omega) \sum_{{\bf k}''_\parallel \alpha''}  \delta U_{{\bf k}_\parallel \alpha, {\bf k}''_\parallel \alpha''}\delta U_{{\bf k}''_\parallel \alpha'', {\bf k}'_\parallel \alpha'}\, G^{(0)}_{{\bf k}'_\parallel \alpha'}(i\omega),
\end{align}
where
\begin{equation}
\delta U_{{\bf k}_\parallel \alpha, {\bf k}'_\parallel \alpha'} \equiv\frac{1}{A}\sum_{k'_z} U_{{\bf k_{\parallel}-k'_{\parallel}}, k'_z}\langle u_{{\bf k_{\parallel}}\alpha}|e^{i  k'_z z }|u_{{\bf k'_{\parallel}}\alpha'} \rangle.
\end{equation}
Next, we average over different realizations of the random potential.
Using $\overline{U_{{\bf k}}}=0$ and $\overline{U_{{\bf k}_\parallel k_z} U_{{\bf k}'_\parallel k'_z}}=(g^{\rm imp}_{{\bf k}_\parallel k_z})^2 \delta_{{\bf k}_\parallel, -{\bf k}'_\parallel} \delta_{k_z,-k'_z}$,
we arrive at
\begin{equation}
\label{eq:av}
\overline{G_{{\bf k}_\parallel \alpha, {\bf k}'_\parallel \alpha'}} (i\omega) =G^{(0)}_{{\bf k}_\parallel \alpha}(i\omega)\, \delta_{\alpha \alpha'} \delta_{{\bf k}_\parallel, {\bf k}'_\parallel}
+ \delta_{{\bf k}_\parallel, {\bf k}'_\parallel}  G^{(0)}_{{\bf k}_\parallel \alpha} (i\omega) G^{(0)}_{{\bf k}_\parallel \alpha'}(i\omega) \sum_{{\bf k}'' \alpha''}  F^{\rm imp}_{{\bf k}_\parallel \alpha, {\bf k}'' \alpha''} (F^{\rm imp})^*_{{\bf k}_\parallel \alpha', {\bf k}'' \alpha''}   G^{(0)}_{{\bf k}''_\parallel \alpha''}(i\omega),
\end{equation}
where 
\begin{equation}
F^{\rm imp}_{{\bf k}_\parallel \alpha, {\bf k}'\alpha'}=g^{\rm imp}_{{\bf k}_\parallel-{\bf k}'_\parallel, k'_z}  \langle u_{{\bf k_{\parallel}}\alpha}|e^{i k'_z z} |u_{{\bf k'_{\parallel}} \alpha'}\rangle/A.
\end{equation}
The underlying assumption in this derivation is that the average over disorder configurations is not altered by the broken translational invariance along the $z$ direction.

From Eq.~(\ref{eq:av}), we may directly read out the electron self-energy.
Since we are interested in the lifetime of a surface electron in band $\alpha$,  we take $\alpha=\alpha'$ hereafter.
Like in the preceding section, we separate the self-energy into an intrasurface and a surface-to-bulk part,
\begin{equation}
\Sigma^{\rm imp}_{{\bf k}_\parallel \alpha}(i\omega)=\Sigma^{\rm imp}_{{\bf k}_\parallel \alpha, {\rm SS}}(i\omega)+\Sigma^{\rm imp}_{{\bf k}_\parallel \alpha,{\rm SB}}(i\omega),
\end{equation}
where
\begin{equation}
\label{eq:apse}
\Sigma^{\rm imp}_{{\bf k}_\parallel \alpha, SB (SS)}(i\omega)=\sum_{{\bf k}'' \alpha''\in B (\alpha''\in S)} |F^{\rm imp}_{{\bf k}_\parallel \alpha, {\bf k}'' \alpha''}|^2 G_{{\bf k}''_\parallel \alpha''}^{(0)}(i\omega)
\end{equation}
By analytically continuing Eq.~(\ref{eq:apse}) and taking the imaginary part of the resulting expression, we arrive at the disorder-induced scattering rate shown in Eq.~(3) of the main text.
 
\subsection{C.~Surface-to-bulk scattering matrix elements: analytical expressions}

The objective of this section is to derive Eq.~(5) of the main text.
The starting point is to solve the $4\times 4$ matrix Hamiltonian ${\cal H}({\bf k}_\parallel,\partial_z)$ in Eq.~(1) of the main text. 
This can be done analytically at ${\bf k_{\parallel}}=0$ (cf. Ref.~[12] in the main text).
In the basis $\{P_1\uparrow,P_1\downarrow,P_2\uparrow,P_2\downarrow\}$, the eigenfunctions read
\begin{eqnarray}
\label{eqk0}
\psi_{\uparrow}^{n,+}=N^{n,+}\left(\begin{matrix} \beta_{+} \eta_{1}^{n,+} f_1^{n,+}\cr0\cr i v_{\perp} f_2^{n,+}\cr0\end{matrix}\right),
\psi_{\uparrow}^{n,-}=N^{n,-}\left(\begin{matrix} \beta_{+} \eta_{2}^{n,-} f_2^{n,-}\cr0\cr i v_{\perp} f_1^{n,-}\cr0\end{matrix}\right),\nonumber\\
\psi_{\downarrow}^{n,+}=N^{n,+}\left(\begin{matrix}0\cr \beta_{+} \eta_{1}^{n,+} f_1^{n,+}\cr 0\cr -i v_{\perp} f_2^{n,+}\end{matrix}\right),
\psi_{\downarrow}^{n,-}=N^{n,-}\left(\begin{matrix}0\cr\beta_{+} \eta_{2}^{n,-} f_2^{n,-}\cr0\cr-i v_{\perp} f_1^{n,-}\end{matrix}\right),
\end{eqnarray} 
where $\pm$ in the superscript denotes positive and negative energy solutions (with eigenenergies $E^{+,n}$ and $E^{-,n}$, respectively), $\uparrow (\downarrow)$ is the spin direction (spin is a good quantum number at $k_\parallel=0$) , $N^{n,\pm}$ are the normalization constants and  $\beta_{\pm}\equiv\beta_{\perp}\pm\gamma_{\perp}$.

The eigenfunctions in Eq.~(\ref{eqk0}) can represent both surface and bulk states.
When referring to bulk states, the index  $n$ represents the $n-$th quantum well state.
States with different $n$ are orthogonal to one another.
In the main text, as well as in the previous sections of this Supplementary Material, we have used $\alpha$ to label different eigenstates.
In case of bulk states, this $\alpha$ may be understood as a ``composite'' label that describes one of the four eigenstates in Eq.~(\ref{eqk0}) (or their appropriate generalizations to finite $k_\parallel$) together with the quantum well label $n$. 
When referring to surface states, the index $n$ is superfluous (it is pinned to a single value) and can be ignored. 

Returning to Eq.~(\ref{eqk0}),  $f^{n,\pm}_{1,2}$ and $\eta^{n,\pm}_{1,2}$ are given by
\begin{align}
f_1^{n,\pm} &=\frac{\sinh(\lambda^{n,\pm}_1 z)}{\sinh(\lambda^{n,\pm}_1 L/2)}-\frac{\sinh(\lambda^{n,\pm}_2 z)}{\sinh(\lambda^{n,\pm}_2 L/2)}\,\,\,;\,\,\,f_2^{n,\pm} =\frac{\cosh(\lambda^{n,\pm}_1 z)}{\cosh(\lambda^{n,\pm}_1 L/2)}-\frac{\cosh(\lambda^{n,\pm}_2 z)}{\cosh(\lambda^{n,\pm}_2 L/2)}\nonumber\\
 \eta^{n,\pm}_1 &=\frac{(\lambda^{n,\pm}_1)^2-(\lambda^{n,\pm}_2)^2}{\lambda^{n,\pm}_1 \coth(\lambda^{n,\pm}_1 L/2)-\lambda^{n,\pm}_2 \coth(\lambda^{n,\pm}_2 L/2)}\,\,\,;\,\,\,\eta^{n,\pm}_2 =\frac{(\lambda^{n,\pm}_1)^2-(\lambda^{n,\pm}_2)^2}{\lambda^{n,\pm}_1 \tanh(\lambda^{n,\pm}_1 L/2)-\lambda^{n,\pm}_2 \tanh(\lambda^{n,\pm}_2 L/2)}.
\end{align}
Here, $\lambda_1$ and $\lambda_2$ are defined via
\begin{equation}
\label{eq:lambda}
\lambda^{n,\pm}_1 = \sqrt{(F^{n,\pm} - \sqrt{R^{n,\pm}})/(2 \beta_{+}\beta_{-})}\,\,\,,\,\,\,\lambda^{n,\pm}_2 = \sqrt{(F^{n,\pm} + \sqrt{R^{n,\pm}})/(2\beta_{+}\beta_{-})},
\end{equation}
where $F^{n,\pm} = v_{\perp}^2 + \beta_{+} (E^{n,\pm}-m)-\beta_{-}(E^{n,\pm}+m)$, $R^{n,\pm} = (F^{n,\pm})^2 + 4 \beta_{+}\beta_{-} ((E^{n,\pm})^2 - m^2)$.
In order to obtain the eigenenergies $E^{n,\pm}$, we solve the following transcendental equations~[12]:
\begin{align}
\label{tran}
\frac{m+\beta_{+} (\lambda_2^{n,+})^2+E^{n,+}}{m+\beta_{+}(\lambda_1^{n,+})^2+E^{n,+}}=\frac{\lambda_2^{n,+} \tanh(\lambda_1^{n,+} L/2)}{\lambda_1^{n,+} \tanh(\lambda_2^{n,+} L/2)}\nonumber\\
\frac{m+\beta_{+} (\lambda_2^{n,-})^2+E^{n,-}}{m+\beta_{+}(\lambda_1^{n,-})^2+E^{n,-}}=\frac{\lambda_2^{n,-} \tanh(\lambda_2^{n,-} L/2)}{\lambda_1^{n,-} \tanh(\lambda_1^{n,-} L/2)}.
\end{align} 
For surface state solutions, $\lambda_1$ and $\lambda_2$ are complex conjugates of each other and their real parts (which are nonzero) give the inverse penetration depth into the bulk.
For bulk state solutions, $\lambda_1$ is purely imaginary and $|\lambda_1|$ plays the role of a quantized momentum along the $z$ direction, while $\lambda_2$ may be either real or purely imaginary.
When $\lambda_2$ is real, it may be understood as the inverse ``healing length'' of the bulk state wave functions, which go from zero at $z=\pm L/2$ to a regular oscillatory (standing wave) behavior deeper into the film. 
If the values of the band parameters are such that the energy bands of the infinite crystal at $k_\parallel=0$ have an absolute minimum at nonzero $k_z$, then for a thin film geometry there will be some solutions with purely imaginary $\lambda_2$. 
For these solutions, $|\lambda_1|$ and $|\lambda_2|$ play the role of the two different momenta that result in the same energy.

The next step is to obtain the eigenfunctions at finite $k_\parallel$, using perturbation theory. 
For the bulk states, we obtain
\begin{align}
\label{eqk}
\Psi_{{\bf k_{\parallel}}1 n}=N_n(\psi_{\uparrow}^{n,+}+ {v_n k_{+}\over\epsilon_{{\bf k_{\parallel}}n}+m_{{\bf k_{\parallel}}n}} \psi_{\downarrow}^{n,-})\nonumber\\
\Psi_{{\bf k_{\parallel}}2 n}=N_n (\psi_{\downarrow}^{n,+}+ {-v_n k_{-}\over\epsilon_{{\bf k_{\parallel}}n}+m_{{\bf k_{\parallel}}n}} \psi_{\uparrow}^{n,-})\nonumber\\
\Psi_{{\bf k_{\parallel}}3 n}=N_n (\psi_{\uparrow}^{n,-}+ {v_n^{\ast} k_{+}\over\epsilon_{{\bf k_{\parallel}}n}+m_{{\bf k_{\parallel}}n}} \psi_{\downarrow}^{n,+})\nonumber\\
\Psi_{{\bf k_{\parallel}}4 n}=N_n (\psi_{\downarrow}^{n,-}+ {-v_n^{\ast}k_{-}\over\epsilon_{{\bf k_{\parallel}}n}+m_{{\bf k_{\parallel}}n}} \psi_{\uparrow}^{n,+}),
\end{align}  
where ${(1,2)}$ and ${(3,4)}$  correspond to two degenerate conduction and valence bands, respectively (not to be confused with the labels 1 and 2 that appear in $\lambda$ and $\eta$),  $N_n$ is the normalization constant, $k_{\pm}\equiv k_{x}\pm k_{y}$,  
\begin{equation}
v_n\equiv v_{\parallel}\langle\psi_{\uparrow}^{n,+}|\sigma^x|\psi_{\downarrow}^{n,-}\rangle
\end{equation}
is an effective in-plane Dirac velocity, 
\begin{equation}
\epsilon_{{\bf k_{\parallel}}n} \equiv\sqrt{m_{{\bf k_{\parallel}}n}^2+|v_n|^2{\bf k_{\parallel}}^2}
\end{equation}
is an effective band energy and
\begin{equation}
m_{{\bf k_{\parallel}}n}\equiv\frac{E^{n,+}-E^{n,-}}{2}-\frac{\beta_\parallel k_\parallel^2}{2} (\psi_{\uparrow}^{n,+}|\sigma^z|\psi_{\uparrow}^{n,+}\rangle-\langle\psi_{\downarrow}^{n,-}|\sigma^z|\psi_{\downarrow}^{n,-}\rangle)
\end{equation}
is an effective Dirac mass.
For the band parameters of Bi$_2$Se$_3$ and related materials, we find that $v_n$ is {\em purely imaginary}. 

Equation~(\ref{eqk}) has been obtained from an effective Hamiltonian computed to leading order in $k_\parallel$; hence the above results are valid only up to terms of order $k_\parallel^2$.
Besides, we have neglected the matrix elements of the perturbation mixing {\em different} quantum well states (i.e. $n$ and $n'\neq n$). 
We have verified numerically that the thicker the film is, the better this approximation holds.
Ultimately, for a film with infinite thickness, the quantum well state index becomes a standing wave involving momenta $\pm k_z$, and it is clear that matrix elements connecting $\pm k_z$ with $\pm k_z'$ are nonzero only if $k_z=k_z'$.
Our approximation, justifiable except for the thinnest films,  is needed in order to make analytical progress in the rest of this section.
However, the numerical results presented in the main text do not rely on it because in those cases we have used the exact solutions of ${\cal H}({\bf k}_\parallel, z)$.

Next, let us find the surface eigenfunctions at finite $k_\parallel$. 
The surface wave functions in Eq.~(\ref{eqk0}) have nonzero projections onto both surfaces of the film. 
However, for all but ultrathin films, the surface eigenfunctions are practically degenerate at $k_\parallel=0$.
This allows us to choose linear combinations of degenerate surface wavefunctions such that the resulting wave functions have all their weight on only one surface of the film. 
For example, in Eq.~(\ref{eqk0}), $\psi_{\uparrow(\downarrow)}^{+}+\psi_{\uparrow(\downarrow)}^{-}$ is localized on the top surface, whereas $\psi_{\uparrow(\downarrow)}^{+}-\psi_{\uparrow(\downarrow)}^{-}$ is localized on the bottom surface (recall that, for surface states, we suppress the index $n$ because it is superfluous). 
From here on, we will concentrate on the surface states localized on the top surface.
At finite $k_\parallel$, the wave function for the surface state localized on the top surface reads
\begin{align}
\label{eqns}
\Psi_{{\bf k_{\parallel}}c}^S=N_{s}\left(\begin{matrix}\beta_{+} \lambda_s e^{-i \theta_{{\bf k_{\parallel}}}}\cr i \beta_{+} \lambda_s \cr i v_{\perp} e^{-i \theta_{{\bf k_{\parallel}}}}\cr v_{\perp} \end{matrix}\right) f_s\,\,\,,\,\,\,
\Psi_{{\bf k_{\parallel}}v}^S=N_{s}\left(\begin{matrix}\beta_{+} \lambda_s e^{-i \theta_{{\bf k_{\parallel}}}}\cr -i \beta_{+} \lambda_s \cr i v_{\perp} e^{-i \theta_{{\bf k_{\parallel}}}}\cr -v_{\perp}\end{matrix}\right) f_s,
\end{align}
where $f_s=(f_1+f_{2})$, $\tan(\theta_{{\bf k_{\parallel}}})=k_y/k_x$, $N_s$ is the normalization constant and $c (v)$ labels the upper (lower) Dirac cone.
In the large $L$ limit, we have $\lambda_s=\lambda_1+\lambda_2=v_{\perp}/\sqrt{\beta_+\beta_-}$. 
Here, we have omitted the superscript $\pm$ in $\lambda_1$ and $\lambda_2$, because $\lambda_{1 (2)}^+=\lambda_{1 (2)}^-$ as can be seen from Eq.~(\ref{tran}) by recognizing that all surface states are degenerate at $k_\parallel=0$.
 
Having obtained the eigenfunctions of ${\cal H}({\bf k}_\parallel, z)$, we are ready to compute surface-to-bulk scattering matrix elements.
To be explicit, we consider an electron on surface band $c$ and calculate the corresponding matrix elements for scattering into the bulk.
We arrive at
\begin{align}
\label{eq:F}
F_{{\bf k}_\parallel c, {\bf k}' 1n'} &\equiv\langle \Psi_{{\bf k_{\parallel}}c}^S|e^{i k'_z z}| \Psi_{{\bf k'_{\parallel}}1n'}^{}\rangle=N_s N_{n'}\left(e^{i \theta_{{\bf k_{\parallel}}}} A_{k'_z,n'}-i {v_{n'} k'_{+}\over \epsilon_{{\bf k}'_{\parallel} n'}+m_{{\bf k}'_{\parallel} n'}}B_{k'_z,n'} \right)\nonumber\\
F_{{\bf k}_\parallel c, {\bf k}' 2n'} &\equiv\langle \Psi_{{\bf k_{\parallel}}c}^S|e^{i k'_z z}| \Psi_{{\bf k'_{\parallel}}2n'}^{}\rangle=N_s N_{n'}\left( {-v_{n'} k'_-\over \epsilon_{{\bf k}'_{\parallel}n'}+m_{{\bf k}'_{\parallel}n'}} e^{i \theta_{\bf k_{\parallel}}}B_{k'_z,n'}-i A_{k'_z,n'}\right)\nonumber\\
F_{{\bf k}_\parallel c, {\bf k}' 3n'} &\equiv\langle \Psi_{{\bf k_{\parallel}}c}^S|e^{i k'_z z}| \Psi_{{\bf k'_{\parallel}}3n'}^{}\rangle=N_s N_{n'}\left(e^{i \theta_{\bf k_{\parallel}}}B_{k'_z,n'}-i {v_{n'}^{\ast} k'_+\over \epsilon_{{\bf k}'_{\parallel}n'}+m_{{\bf k}'_{\parallel} n'}}A_{k'_z,n'}\right)\nonumber\\
F_{{\bf k}_\parallel c, {\bf k}' 4n'} &\equiv\langle \Psi_{{\bf k_{\parallel}}c}^S|e^{i k'_z z}| \Psi_{{\bf k'_{\parallel}}4n'}^{}\rangle=N_s N_{n'}\left({ -v_{n'}^{\ast} k'_-\over \epsilon_{{\bf k}'_{\parallel}n'}+m_{{\bf k}'_{\parallel}n'} }e^{i \theta_{\bf k_{\parallel}}}A_{k'_z,n'}-i B_{k'_z,n'}\right),
\end{align}
where 
\begin{align}
\label{form}
A_{k'_z,n'} &=N^{n',+}\left[\beta_+ ^2\eta_1^{n',+} \lambda_s \int (f_1^{n',+})^{\ast} f_s \,e^{i k'_z z} dz + v_{\perp}^2 \int (f_2^{n',+})^{\ast} f_s\, e^{i k'_z z} dz \right]\nonumber\\
B_{k'_z,n'} &=N^{n',-}\left[\beta_+^2 \eta_2^{n',-} \lambda_s \int (f_2^{n',-})^{\ast} f_s \,e^{i k'_z z} dz +v_{\perp}^2 \int (f_1^{n',-})^{\ast} f_s \,e^{i k'_z z} dz \right]
\end{align}
and the $z-$integration runs from $-L/2$ to $+L/2$.
The first two lines in Eq.~(\ref{eq:F}) describe transitions from the upper Dirac cone on the surface to a bulk conduction band.
The last two lines describe transitions from the upper Dirac cone on the surface to a bulk valence band.
Typically the latter transitions do not enter directly in physical quantities (because the largest phonon frequency is small compared to the bulk bandgap in most topological materials); however, they will be useful below in order to derive Eq.~(5) of the main text.

The relation between $F$ in Eq.~(\ref{eq:F}) and the expression from Eq.~(4) in the main text is $\sum F=F^{\rm imp (ph)}/g^{\rm imp (ph)}$, where the sum is over the two degenerate bulk bands.
The appearance of this sum is natural, because each bulk state is doubly degenerate (i.e. $1 n'$ and $2 n'$ are degenerate, as are $3 n'$ and $4 n'$).
For that reason, when computing the SB scattering rate of a surface electron in band $c$, the matrix element that matters is $|F_{{\bf k}_\parallel c, {\bf k}' 1n'}|^2+|F_{{\bf k}_\parallel c, {\bf k}' 2n'}|^2$, or $|F_{{\bf k}_\parallel c, {\bf k}' 3n'}|^2+|F_{{\bf k}_\parallel c, {\bf k}' 4n'}|^2$ (although, as mentioned above, the latter does not contribute in most toplogical materials). 
With this in mind, we concentrate on the following expression:
\begin{align}
\label{mat}
&\left|F_{{\bf k}_\parallel c, {\bf k}' 1n'}\right|^2+\left|F_{{\bf k}_\parallel c, {\bf k}' 2n'}\right|^2\nonumber\\
&= N_s^2 N_{n'}^2 \left(2|A_{k'_z,n'}|^2+2 |B_{k'_z,n'}|^2 {|v_{n'}|^2{ {\bf k}'_{\parallel}}^2\over(\epsilon_{{\bf k}'_{\parallel}n'}+m_{{\bf k}'_{\parallel}n'})^2} +|v_{n'}| {(B_{k'_z,n'}^{\ast} A_{k'_z,n'}+{\rm c.c.})(k'_- e^{i \theta_{\bf k_{\parallel}}}+ {\rm c.c.}) \over(\epsilon_{{\bf k}'_\parallel n'}+m_{{\bf k}'_{\parallel}n'})}\right),
\end{align}
where in the last term we have used $v_{n'}=i |v_{n'}|$.
Similarly, we can calculate
\begin{align}
\label{mat2}
&\left|F_{{\bf k}_\parallel c, {\bf k}' 3n'}\right|^2+\left|F_{{\bf k}_\parallel c, {\bf k}' 4n'}\right|^2\nonumber\\
&= N_s^2 N_{n'}^2 \left(2|B_{k'_z,n'}|^2+2|A_{k'_z,n'}|^2 {|v_{n'}|^2 {{\bf k}'_{\parallel}}^2\over(\epsilon_{{\bf k}'_{\parallel} n'}+m_{{\bf k}'_{\parallel} n'})^2} - |v_{n'}| {( A_{k'_z,n'} B_{k'_z,n'}^{\ast} +{\rm c.c.})(k'_- e^{i \theta_{\bf k_{\parallel}}}+ {\rm c.c.}) \over(\epsilon_{{\bf k'}_{\parallel}n'}+m_{{\bf k'}_{\parallel}n'})}\right).
\end{align}
The differences and similarities between Eqs.~(\ref{mat}) and~(\ref{mat2}) are suggestive of a simple underlying mathematical structure.
In order to uncover it, we write
\begin{align}
\label{feq}
&|F_{{\bf k}_\parallel c, {\bf k}' 1n'}|^2+|F_{{\bf k}_\parallel c, {\bf k}' 2n'}|^2\nonumber\\
&={1 \over 2}\left(|F_{{\bf k}_\parallel c, {\bf k}' 1n'}|^2+|F_{{\bf k}_\parallel c, {\bf k}' 2n'}|^2+|F_{{\bf k}_\parallel c, {\bf k}' 3n'}|^2+|F_{{\bf k}_\parallel c, {\bf k}' 4n'}|^2\right)+{1\over 2}\left(|F_{{\bf k}_\parallel c, {\bf k}' 1n'}|^2+|F_{{\bf k}_\parallel c, {\bf k}' 2n'}|^2-|F_{{\bf k}_\parallel c, {\bf k}' 3n'}|^2-|F_{{\bf k}_\parallel c, {\bf k}' 4n'}|^2\right)\nonumber\\
&=N_s^2 (|A_{k'_z,n'}|^2+|B_{k'_z,n'}|^2)  \left[1+  \frac{|A_{k'_z,n'}|^2-|B_{k'_z,n'}|^2}{|A_{k'_z,n'}|^2+|B_{k'_z,n'}|^2}\frac{ {(1- {|v_{n'}|^2 {{\bf k}'_{\parallel}}^2\over(\epsilon_{{\bf k}'_{\parallel}n'}+m_{{\bf k}'_{\parallel}n'})^2})}}{{(1+ {|v_{n'}|^2 {{\bf k}'_{\parallel}}^2\over(\epsilon_{{\bf k}'_{\parallel}n'}+m_{{\bf k}'_{\parallel}n'})^2})}}\right.\nonumber\\
&\left.~~~~~~~~~~~~~~~~~~~~~~~~~~~~~~~~~+{ 2 |v_{n'}| (A_{k'_z,n'} B_{k'_z,n'}^* +{\rm c.c.})\over (|A_{k'_z,n'}|^2+|B_{k'_z,n'}|^2) } \frac{ k'_x \cos(\theta_{\bf k_{\parallel}})+k'_y \sin(\theta_{\bf k_{\parallel}} )}{(1 + {|v_{n'}|^2 {k'_{\parallel}}^2\over(\epsilon_{{\bf k}'_{\parallel}n'}+m_{{\bf k}'_{\parallel}n'})^2}) (\epsilon_{{\bf k}'_{\parallel}n'}+m_{{\bf k}'_{\parallel}n'})}\right]
\end{align}
and
\begin{align}
\label{eqval}
& |F^{\rm imp (ph)}_{{\bf k}_\parallel c, {\bf k}' 3n'}|^2+|F^{\rm imp (ph)}_{{\bf k}_\parallel c, {\bf k}' 4n'}|^2\nonumber\\
&={1 \over 2}\left(|F_{{\bf k}_\parallel c, {\bf k}' 3n'}|^2+|F_{{\bf k}_\parallel c, {\bf k}' 4n'}|^2+|F_{{\bf k}_\parallel c, {\bf k}' 1n'}|^2+|F_{{\bf k}_\parallel c, {\bf k}' 2n'}|^2\right)+{1\over 2}\left(|F_{{\bf k}_\parallel c, {\bf k}' 3n'}|^2+|F_{{\bf k}_\parallel c, {\bf k}' 4n'}|^2-|F_{{\bf k}_\parallel c, {\bf k}' 1n'}|^2-|F_{{\bf k}_\parallel c, {\bf k}' 2n'}|^2\right)\nonumber\\
&=N_s^2 (|A_{k'_z,n'}|^2+|B_{k'_z,n'}|^2)  \left[1 -  \frac{|A_{k'_z,n'}|^2-|B_{k'_z,n'}|^2}{|A_{k'_z,n'}|^2+|B_{k'_z,n'}|^2}\frac{ {(1- {|v_{n'}|^2 {{\bf k}'_{\parallel}}^2\over(\epsilon_{{\bf k}'_{\parallel}n'}+m_{{\bf k}'_{\parallel}n'})^2})}}{{(1+ {|v_{n'}|^2 {{\bf k}'_{\parallel}}^2\over(\epsilon_{{\bf k}'_{\parallel}n'}+m_{{\bf k}'_{\parallel}n'})^2})}}\right.\nonumber\\
&\left.~~~~~~~~~~~~~~~~~~~~~~~~~~~~~~~~~-{ 2 |v_{n'}| (A_{k'_z,n'} B_{k'_z,n'}^* +{\rm c.c.})\over (|A_{k'_z,n'}|^2+|B_{k'_z,n'}|^2) } \frac{ k'_x \cos(\theta_{\bf k_{\parallel}})+k'_y \sin(\theta_{\bf k_{\parallel}} )}{(1 + {|v_{n'}|^2 {k'_{\parallel}}^2\over(\epsilon_{{\bf k}'_{\parallel}n'}+m_{{\bf k}'_{\parallel}n'})^2}) (\epsilon_{{\bf k}'_{\parallel}n'}+m_{{\bf k}'_{\parallel}n'})}\right]
\end{align}
The last two terms in Eqs.~(\ref{feq}) and (\ref{eqval}) differ in sign; it turns out that this difference can be ascribed to the change in the expectation value of certain operators when going from the bulk conduction band to the bulk valence band.
We guess the answer to be of the form
\begin{align}
\label{ins}
& \frac{(|A_{k'_z,n'}|^2-|B_{k'_z,n'}|^2)}{(|A_{k'_z,n'}|^2+|B_{k'_z,n'}|^2)}\frac{ {(1- {|v_{n'}|^2 {k'_{\parallel}}^2\over(\epsilon_{{\bf k}'_{\parallel}n'}+m_{{\bf k}'_{\parallel}n'})^2})}}{{(1+ {|v_{n'}|^2 {k'_{\parallel}}^2\over(\epsilon_{{\bf k}'_{\parallel}n'}+m_{{\bf k}'_{\parallel}n'})^2})}}=\xi_{k'_z,n'}  
\langle\tau^{z}\rangle^S_{{\bf k_{\parallel}}c} \langle\tau^{z}\rangle^{B}_{{\bf k}'_{\parallel}1n'}\\
\label{ins2}
&{ 2 |v_{n'}| (A_{k'_z,n'} B_{k'_z,n'}^{\ast}  +{\rm c.c.})\over (|A_{k'_z,n'}|^2+|B_{k'_z,n'}|^2) } \frac{ k'_x \cos(\theta_{\bf k_{\parallel}})+k'_y \sin(\theta_{\bf k_{\parallel}} )}{(1 + {|v_{n'}|^2 {k'_{\parallel}}^2\over(\epsilon_{{\bf k}'_{\parallel}n'}+m_{{\bf k}'_{\parallel}n'})^2}) (\epsilon_{{\bf k}'_{\parallel}n'}+m_{{\bf k}'_{\parallel}n'})}=\xi'_{k'_z,n'} \sum_{i\in \{x,y,z\}}\langle \sigma^i\tau^x\rangle^S_{{\bf k_{\parallel}}c} \langle \sigma^i \tau^x\rangle^{B}_{{\bf k'_{\parallel}}1n'}.
\end{align}
Next, we will justify our guess. 
First, let us discuss Eq.~(\ref{ins}). 
On the right hand side (rhs),  $\langle\tau^{z}\rangle^S_{{\bf k_{\parallel}}c}$ is given by 
\begin{equation}
\label{eq:taus}
\langle\tau^{z}\rangle^S_{{\bf k_{\parallel}}c}=\langle u_{{\bf k}_\parallel c\in S}|\tau^z|u_{{\bf k}_\parallel c\in S}\rangle/A=\langle \Psi^S_{{\bf k}_\parallel c}|\tau^z|\Psi^S_{{\bf k}_\parallel c}\rangle=2 N_s^2 (\beta_+^2\lambda_s^2-v_{\perp}^2) \int |f_{s}(z)|^2 dz=\frac{(\beta_+^2\lambda_s^2-v_{\perp}^2)}{(\beta_+^2\lambda_s^2+v_{\perp}^2)}=\frac{\gamma_{\perp}}{\beta_{\perp}},
\end{equation}
where we have used Eq.~(\ref{eqns}) and assumed $\gamma_{\perp}\le\beta_{\perp}$.
For $\gamma_{\perp}>\beta_{\perp}$, the surface Dirac point would buried be into the $k_\parallel=0$ bulk band and $\langle\tau^{z}\rangle^S_{{\bf k_{\parallel}}c}=\beta_\perp/\gamma_\perp$.
In this work, the case of interest is $\gamma_{\perp}\le\beta_{\perp}$.
Note that $\langle\tau^{z}\rangle^S_{{\bf k_{\parallel}}c}$ is independent of $k_\parallel$.
While this result has been derived perturbatively for small $k_\parallel$, we have checked (by solving ${\cal H}({\bf k}_\parallel, z)$ numerically) that the dependence of $\langle\tau^{z}\rangle^S_{{\bf k_{\parallel}}c}$ on $k_\parallel$ is rather weak for a wide interval of $k_\parallel$. 

Similarly, $\langle\tau^{z}\rangle^{B}_{{\bf k'_{\parallel}}1n'}$ on the rhs of Eq.~(\ref{ins}) is given by
\begin{equation}
\langle\tau^{z}\rangle^{B}_{{\bf k'_{\parallel}}1n'}=\frac{ {(1- {|v_{n'}|^2 {k'_{\parallel}}^2\over(\epsilon_{{\bf k}'_{\parallel}n'}+m_{{\bf k}'_{\parallel}n'})^2})}}{{(1+ {|v_{n'}|^2 {k'_{\parallel}}^2\over(\epsilon_{{\bf k}'_{\parallel}n'}+m_{{\bf k}'_{\parallel}n'})^2})}}\frac{m_{\lambda_{1} n'}}{\epsilon_{\lambda_1 n'}}\simeq \frac{M_{{\bf k'_{\parallel}},\lambda_1}}{E_{{\bf k'_{\parallel}},1 n'}},
\label{eq:btz}
\end{equation}
where  $m_{\lambda_1 n}=m-\beta_\perp |\lambda_1^{n,+}|^2$, $\epsilon_{\lambda_{1}n}=\sqrt{m_{\lambda_{1}n}^2+v_{\perp}^2 |\lambda_{1}^{n,+}|^2}$, $M_{{\bf k_{\parallel}},\lambda_{1}}=m_{\lambda_{1 }n}-\beta_\parallel k_\parallel^2$ and $E_{{\bf k_{\parallel}},1 n'}=\sqrt{v_\perp |\lambda_{1}^{n,+}|^2 + v_\parallel k_\parallel^2+M^2_{{\bf k}_\parallel,\lambda_{1}}}$. 
In the derivation of Eq.~(\ref{eq:btz}), we have  used Eq.~(\ref{eqk}) and have considered the case of real $\lambda_2$ and purely imaginary $\lambda_1$.
This case comprises many situations of interest and serves to illustrate our calculation.
In addition, we have used the following relations: $N^{n,+} v_\perp = N^{n,-} \beta_+ \eta_2^{n,-}$ and $\eta_1^{n,+}\eta_2^{n,+}=\eta_1^{n,-}\eta_2^{n,-}=v_\perp^2/\beta_+^2$.
The latter relation is generally valid, whereas the former relation relies on $\lambda_1^{n,+}\simeq \lambda_1^{n,-}$, which is satisfied even when particle-hole symmetry is strongly broken.
The last equality in Eq.~(\ref{eq:btz}) has been derived for bulk states close to the bulk band edge and close to the Brillouin zone center.
This is a relevant situation for the study of bulk surface coupling.
It is also worth noting that $\langle\tau^{z}\rangle^{B}_{{\bf k'_{\parallel}}1n'}$ is independent of the presence or absence of particle-hole symmetry (as expected from the matrix structure of Eq.~(1) in the main text).
Finally, we remark that $\langle\tau^{z}\rangle^{B}_{{\bf k'_{\parallel}}1n'}=\langle\tau^{z}\rangle^{B}_{{\bf k'_{\parallel}}2n'}=-\langle\tau^{z}\rangle^{B}_{{\bf k'_{\parallel}}3n'}=-\langle\tau^{z}\rangle^{B}_{{\bf k'_{\parallel}}4n'}$; this will be useful below.

\begin{figure}
\rotatebox{0}{\includegraphics*[width=\linewidth]{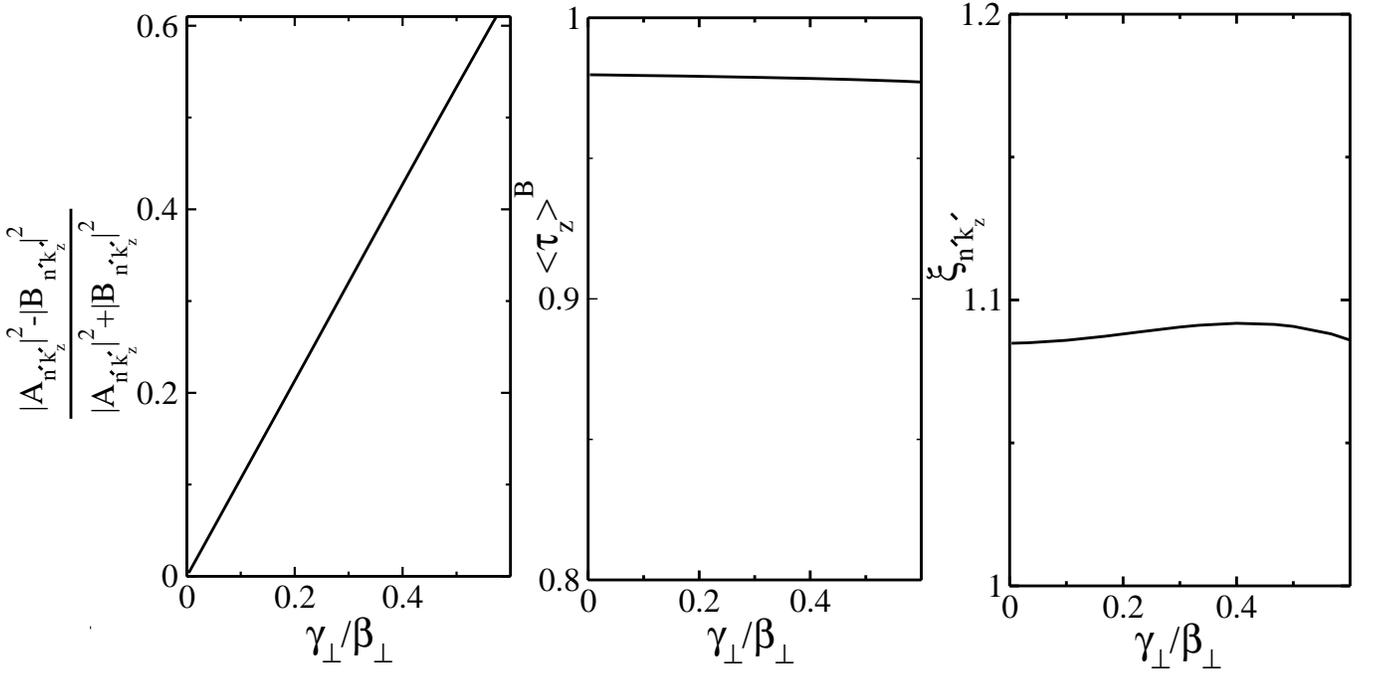}}
\caption{(Left) Plot of lhs of Eq. (\ref{ins}) (at $k_\parallel=0$) as a function of $\gamma_{\perp}/\beta_{\perp}$, for fixed $n'$ and $k_z'$. (Center) $\langle\tau^{z}\rangle^B_{{\bf k'}n'}$  as a function of $\gamma_{\perp}/\beta_{\perp}$ ($\beta_\perp$ is kept fixed, while $\gamma_\perp$ is varied). (Right) $\xi_{n' k_z'}$ as a function of $\gamma_{\perp}/\beta_{\perp}$ ($\beta_\perp$ is kept fixed, while $\gamma_\perp$ is varied).}  
 \label{fig:apn1}
\end{figure}

In order to justify Eq.~(\ref{ins}), we present the following three statements:

(1) For some fixed $n'$ and $k'_z$, we find numerically that the lhs of Eq.~({\ref{ins}}) varies linearly with $\gamma_\perp$, as shown in Fig.~\ref{fig:apn1}a.
In particular, the lhs changes sign when $\gamma_\perp$ changes sign.
In view of Eq.~(\ref{eq:taus}), this suggests that the lhs of Eq.~({\ref{ins}}) is proportional to $\langle\tau^{z}\rangle^S_{{\bf k_{\parallel}}c}$.
Note that the presence of $\langle\tau^{z}\rangle^B_{{\bf k_{\parallel}'}c}$ and $\xi_{k'_z,n'}$ on the rhs does not spoil our guess because both are largely independent of $\gamma_{\perp}/\beta_{\perp}$ as shown in Fig.~\ref{fig:apn1}b and~\ref{fig:apn1}c .

(2) The proportionality of the left hand side (lhs) of Eq.~({\ref{ins}}) to $\langle\tau^{z}\rangle^{B}_{{\bf k'_{\parallel}}1n'}$ is partly justified on the basis that the second terms in the right hand side of Eqs.~(\ref{feq}) and (\ref{eqval}) have opposite signs.
Given that Eq.~(\ref{feq}) describes a transition to bulk conduction band whereas Eq.~(\ref{eqval}) describes transitions to a bulk valence band, the change in sign between the two can be ascribed to the fact that  $\langle\tau^{z}\rangle^{B}$ changes sign from the conduction to the valence band. 
Another way to see that the lhs of Eq.~({\ref{ins}}) is proportional to $\langle\tau^{z}\rangle^{B}_{{\bf k'_{\parallel}}1n'}$ is to observe the equal $k_\parallel$-dependence of 
Eqs.~(\ref{ins}) and (\ref{eq:btz}).
Note that the presence of $\langle\tau^{z}\rangle^S_{{\bf k_{\parallel}'}c}$ and $\xi_{k'_z,n'}$ on the rhs does not spoil our guess because both are independent of $k_\parallel$.

(3) Let us suppose for a moment that we take periodic boundary conditions for the bulk states. This would yield the following bulk eigenstates (taking $m>0$ by convention):
\begin{eqnarray}
\label{eqbp}
\Psi^{}_{{\bf k_{\parallel}},1k_z} =N\left(\begin{matrix}1\cr0\cr \frac{v_\perp k_z}{\epsilon_{\bf k}+M_{\bf k}}\cr \frac{v_\parallel k_{+}}{\epsilon_{\bf k}+M_{\bf k}}\end{matrix}\right)e^{i k_z z};\,\,
\Psi^{}_{{\bf k_{\parallel}},2k_z} =N\left(\begin{matrix}0\cr1\cr \frac{v_\parallel k_{-}}{\epsilon_{\bf k}+M_{\bf k}}\cr \frac{-v_\perp k_z}{\epsilon_{\bf k}+M_{\bf k}} \end{matrix}\right)e^{i k_z z};\,\,
\Psi^{}_{{\bf k_{\parallel}},3k_z} =N\left(\begin{matrix}\frac{-v_\perp k_{z}}{\epsilon_{\bf k}+M_{\bf k}}\cr \frac{-v_\parallel k_+}{\epsilon_{\bf k}+M_{\bf k}} \cr1 \cr0 \end{matrix}\right)e^{i k_z z};\,\,
\Psi^{}_{{\bf k_{\parallel}},4k_z} =N\left(\begin{matrix} \frac{-v_\parallel k_-}{\epsilon_{\bf k}+M_{\bf k}} \cr  \frac{v_\perp k_{z}}{\epsilon_{\bf k}+m_{\bf k}}  \cr 0 \cr 1\end{matrix}\right) e^{i k_z z},\nonumber\\
\end{eqnarray} 
where ${\bf k}=({\bf k_{\parallel}},k_z)$. 
If we combine these bulk states with the surface states of Eq.~(\ref{eqns}), we obtain 
\begin{equation}
\label{eq:bulk}
|F_{{\bf k}_\parallel c, {\bf k}' 1}|^2+|F_{{\bf k}_\parallel c, {\bf k}' 2}|^2\propto 1+ \langle\tau^{z}\rangle^S_{{\bf k_{\parallel}}c} \langle\tau^{z}\rangle_{{\bf k'} 1}^B+ \sum_{i\in \{x,y,z\}}\langle \sigma^i\tau^x\rangle^S_{{\bf k_{\parallel}}c} \langle \sigma^i \tau^x\rangle^{B}_{{\bf k'}1 }
\end{equation} 
by {\em direct} calculation.
Comparing this expression with Eqs.~(\ref{ins}) and (\ref{ins2}), we conclude that  the appearance of  $\xi_{n',k'_z}$ and $\xi'_{n',k'_z}$ is a manifestation of the open boundary conditions for the bulk states, which complicates the analytical treatment of the matrix elements.
However, Eq.~(\ref{eq:bulk}) is highly suggestive that the structure unveiled in Eqs.~(\ref{ins}) and (\ref{ins2}) is correct.

Next, let us discuss Eq.~(\ref{ins2}).
In the rhs of the equation,  $\langle \sigma^i\tau^x\rangle^S_{{\bf k_{\parallel}}c}$ ($i=x,y,z$) are obtained as
\begin{equation}
\label{subeq1}
\langle \sigma^x\tau^x\rangle^S_{{\bf k_{\parallel}}c} = \frac{\sqrt{\beta_+\beta_-}}{\beta_{\perp}} \cos(\theta_{{\bf k}_{\parallel}})\,\,\,;\,\,\,\langle \sigma^y\tau^x\rangle^S_{{\bf k_{\parallel}}c} =\frac{\sqrt{\beta_+\beta_-}}{\beta_{\perp}} \sin(\theta_{{\bf k}_{\parallel}})\,\,\,; \,\,\,\langle \sigma^z\tau^x\rangle^S_{{\bf k_{\parallel}}c} =0,
\end{equation}
where we have used $\lambda_s=v_{\perp}/\sqrt{\beta_+\beta_-}$.  
The counterparts for the bulk states are 
\begin{equation}
\label{subeq2}
\langle \sigma^x\tau^x\rangle^{B}_{{\bf k'_{\parallel}}1n'} \simeq \frac{2 |v_{n'}| k'_x}{(\epsilon_{{\bf k}'_{\parallel}n'}+m_{{\bf k}'_{\parallel}n'})}\frac{1}{(1+ {|v_{n'}|^2 {{\bf k}'_{\parallel}}^2\over(\epsilon_{{\bf k}'_{\parallel}n'}+m_{{\bf k}'_{\parallel}n'})^2})}\,\,\,;\,\,\,
\langle \sigma^y\tau^x\rangle^{B}_{{\bf k'_{\parallel}}1n'}\simeq\frac{2 |v_{n'}| k'_y}{ \epsilon_{{\bf k}'_{\parallel}n'}+m_{{\bf k}'_{\parallel}n'}}\frac{1}{(1+ {|v_{n'}|^2 {{\bf k}'_{\parallel}}^2\over(\epsilon_{{\bf k}'_{\parallel}n'}+m_{{\bf k}'_{\parallel}n'})^2})}\,\,\,;\,\,\,
\langle \sigma^z\tau^x\rangle^{B}_{{\bf k'_{\parallel}}1n'}\simeq 0,
\end{equation}
where similar approximations and assumptions as in Eq.~(\ref{eq:btz}) have been used.

The justification of Eq.~(\ref{ins2}) can be done along the same lines as that of Eq.~(\ref{ins}).
In particular, the statement \# 3 made above holds for Eq.~(\ref{ins2}) too, as is apparent from Eq.~(\ref{eq:bulk}).
Another evidence in favor of Eq.~(\ref{ins2}) is that the dependence of its rhs on $k_\parallel$ exactly matches with that of its lhs; in order to see this, it suffices to combine Eqs.~(\ref{subeq1}) and (\ref{subeq2}).

Having proven Eqs.~(\ref{ins}) and (\ref{ins2}), the derivation of Eq.~(5) in the main text becomes immediate.
All one has to do is to repeat the preceding calculations for the case of a surface electron in band $v$, and verify that the guess presented above holds.
In order to do that, one must use the following relations:

(i) $\langle\tau^{z}\rangle^S_{{\bf k_{\parallel}}c}=\langle \tau^{z}\rangle^S_{{\bf k_{\parallel}}v}$ (i.e. the expectation value of the orbital pseudospin on the surface is the same for the upper and lower halfs of the Dirac cone),

(ii) $\langle\tau^{z}\rangle^{B}_{{\bf k'_{\parallel}}1n'}=\langle\tau^{z}\rangle^{B}_{{\bf k'_{\parallel}}2n'}=-\langle\tau^{z}\rangle^{B}_{{\bf k'_{\parallel}}3n'}=-\langle\tau^{z}\rangle^{B}_{{\bf k'_{\parallel}}4n'}$  (i.e. the expectation values of the orbital pseudospin on the  $n'$-th bulk conduction and $n'$-th valence band are equal in magnitude and opposite in sign),

(iii) $\langle \sigma^i\tau^x\rangle^S_{{\bf k_{\parallel}}c}=-\langle \sigma^i\tau^x\rangle^S_{{\bf k_{\parallel}}v}$  and $\langle \sigma^i\tau^x\rangle^{B}_{{\bf k'_{\parallel}}1n'}=\langle \sigma^i\tau^x\rangle^{B}_{{\bf k'_{\parallel}}2n'}=-\langle \sigma^i\tau^x\rangle^{B}_{{\bf k'_{\parallel}}3n'}=-\langle \sigma^i\tau^x\rangle^{B}_{{\bf k'_{\parallel}}4n'}$.

All the relative signs in the above relations are crucial for the justification of Eq.~(5) in the main text.
    
\subsection{D.~ Phonon- and disorder-induced surface-to-bulk scattering rate }

The objective of this section is to derive Eqs.~(7) and (9) from the main text.
Let us start from the disorder-induced SB scattering rate for a surface electron located at the Fermi level in band $c$ (i.e. in the upper half of the Dirac cone).
It reads
\begin{align}
\label{eq:gamma_dis}
\Gamma_{{\bf k_{\parallel}}c,SB}^{{\rm imp}}&=2\pi\sum_{{\bf k'}n'} |g^{\rm dis}|^2  (|F_{{\bf k}_\parallel c, {\bf k}' 1n'}|^2+|F_{{\bf k}_\parallel c, {\bf k}' 2n'}|^2 )\delta(\xi_{{\bf k'_{\parallel}} 1n'})\nonumber\\
&\simeq |g^{\rm imp}_{{\rm eff}}|^2   \nu_B(\epsilon) \left(1+\langle\tau^z\rangle^S_{{\bf k}_\parallel c}\langle\langle\tau^z\rangle\rangle^B_{\epsilon_F}\right), 
\end{align}
where 
\begin{equation}
\label{eq:gg}
|g^{\rm imp}_{\rm eff}|^2 = 2\pi N_s^2 |g^{\rm imp}|^2\sum_{k'_z} (|A_{k'_z,n'}|^2+|B_{k'_z,n'}|^2) 
\end{equation}
is an effective coupling constant for electron-disorder scattering,
\begin{equation}
\zeta_{n'} = {\sum_{k'_z}{(|A_{k'_z,n'}|^2+|B_{k'_z,n'}|^2) \zeta_{k'_z n'}}\over \sum_{k'_z} (|A_{k'_z,n'}|^2+|B_{k'_z,n'}|^2)}
\end{equation}
is a weighted average of the function $\zeta_{k_z' n'}$ introduced in Eq.~(\ref{ins}) (see also Fig.~\ref{fig:apn3}), and
\begin{equation}
\langle\langle\tau^z\rangle\rangle^B_{\epsilon_F} =\frac{1}{\nu_B (\epsilon_F)}\frac{1}{V}\sum_{{\bf k}'_\parallel n'} \zeta_{n'}\langle\tau^z\rangle^B_{{\bf k}'_\parallel n'}\delta(\xi_{{\bf k}'_\parallel n'})=\frac{\sum_{{\bf k}'_\parallel n'} \zeta_{n'}\langle\tau^z\rangle^B_{{\bf k}'_\parallel n'}\delta(\xi_{{\bf k}'_\parallel n'})}{\sum_{{\bf k}'_\parallel n'} \delta(\xi_{{\bf k}'_\parallel n'})}
\end{equation}
is a Fermi surface average of $\langle\tau^z\rangle^B$ weighted by $\zeta_{n'}$.
In the first line of Eq.~(\ref{eq:gamma_dis}), we have used the fact (alluded to in the previous section) that the bulk bands $1 n'$ and $2 n'$ are degenerate.
The delta function precludes the bulk  valence bands ($3 n'$ and $4 n'$) from contributing to the sum over intermediate states, because we have assumed that the Fermi level is in the conduction band.
We have also assumed that the disorder potential is independent of momentum.
Then, going from the first to the second line of Eq.~(\ref{eq:gamma_dis}), we have used Eqs.~(\ref{feq}),~(\ref{ins}) and (\ref{ins2}). 
We have also taken advantage of the fact that the energy eigenvalues do not depend on $k_z'$.
In addition, we have recognized that Eq.~(\ref{ins2}) makes a vanishing contribution because positive and negative $k_\parallel$ cancel each other.
Thus, we have arrived at Eq.~(7) in the main text (second line in Eq.~(\ref{eq:gamma_dis})).

The derivation of Eq.~(9) in the main text is essentially identical to the one presented above and thus will not be repeated here.
Note that, for optical phonons, $g^{\rm ph}$ is independent of momentum.
The resulting $g^{\rm ph}_{\rm eff}$ is related to $g^{\rm ph}$ in the same way as $g^{\rm imp}_{\rm eff}$ relates to $g^{\rm imp}$ in Eq.~(\ref{eq:gg}).
Regarding acoustic phonons, Eq.~(9) is still applicable under the quasi-elastic approximation, wherein the phonon frequency is neglected inside the delta function.
This approximation is justified when the Fermi energy measured from the bottom of the bulk band is large compared to the Debye frequency and the temperature.

\begin{figure}
\rotatebox{0}{\includegraphics*[width=\linewidth]{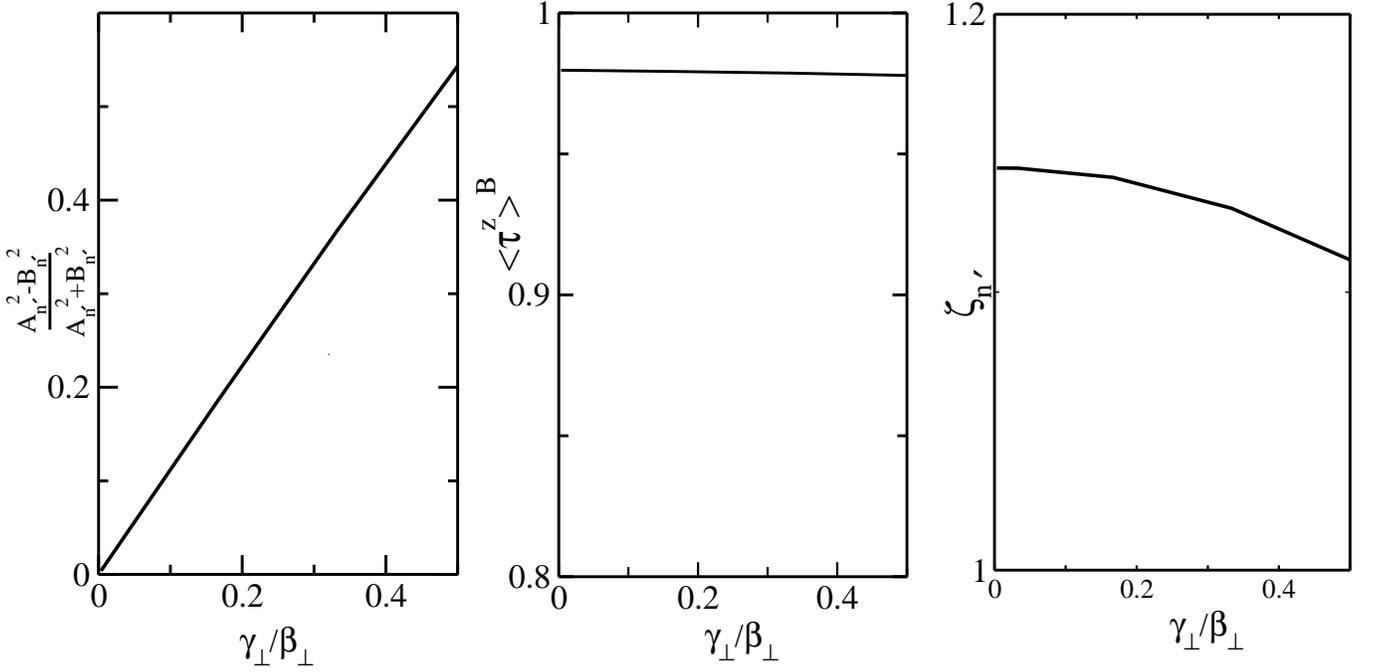}}
\caption{(Left) Plot of $\sum_{k_z'} (|A_{k_z',n'}|^2-|B_{k_z',n'}|^2)/\sum_{k_z'} (|A_{k_z',n'}|^2+|B_{k_z',n'}|^2)$ as a function of $\gamma_{\perp}/\beta_{\perp}$, for a given  $n'$ ($\beta_\perp$ is kept fixed, while $\gamma_\perp$ is varied). (Center) $\langle\tau^{z}\rangle^B$  as a function of $\gamma_{\perp}/\beta_{\perp}$ for a given conduction band ($\beta_\perp$ is kept fixed, while $\gamma_\perp$ is varied).  (Right) $\xi_{n'}$ as a function of $\gamma_{\perp}/\beta_{\perp}$ for a particular $n'$ ($\beta_\perp$ is kept fixed, while $\gamma_\perp$ is varied).} 
 \label{fig:apn3}
\end{figure}

\subsection{E.~Asymmetry on the surface-to-bulk scattering matrix elements between electron- and hole-doped films}

The objective of this section is to provide an example of the asymmetry of the SB scattering rate when the material changes from $n-$doped to $p-$doped.
This effect requires breaking particle-hole symmetry, as otherwise the surface orbital pseudospin would be unpolarized and the SB scattering rate would be identical in electron- and hole-doped systems. 
In Fig.~\ref{fig:apn3}, we have shown that $\zeta_{n'}$ is positive, which means that the SB scattering rate is enhanced if the orbital pseudospin of the bulk and surface states are {\em aligned}.
This point has been emphasized in Fig.~2 of the main text.

Suppose that particle-hole symmetry is broken in such a way that the Dirac point of the surface states is closer to the bulk conduction band than to the bulk valence band.
This means that the bulk and surface orbital pseudospins are parallel to each other  in the conduction band, whereas they are antiparallel to each other in the valence band.
Accordingly, the SB scattering matrix elements should be stronger in an $n-$doped sample than in a $p-$doped sample.
The situation would be opposite if the Dirac point of the surface states were closer to the bulk valence band than to the bulk conduction band.

Figure~\ref{fig:apn2}a illustrates the preceding point. 
We plot only the scattering matrix elements in order to filter out the contribution from the bulk density of states.
We keep the Fermi level in the bulk conduction band but vary the sign of $\gamma_\perp$ so that the surface Dirac point can be (i) close to the bulk conduction or (ii) close to the bulk valence band.
We find the asymmetry in the SB matrix element between cases (i) and (ii) exceeds what would have been expected simply from the relative orientation of the pseudospins.
The reason for this enhanced asymmetry is the proportionality factor of Eq.~(5) in the main text (or Eq.~(\ref{eqval})). 
When the Dirac point is closer to the conduction band, the surface states near the bottom of the conduction band penetrate deeper into the bulk than those near the top of the valence band (cf. Fig.~\ref{fig:apn2}b).
Accordingly, the wave function overlap in the conduction band is stronger in the conduction band, which results in a larger SB scattering matrix element (independently from the relative orientation of the bulk and surface pseudospin).

\begin{figure}
\rotatebox{0}{\includegraphics*[width=\linewidth]{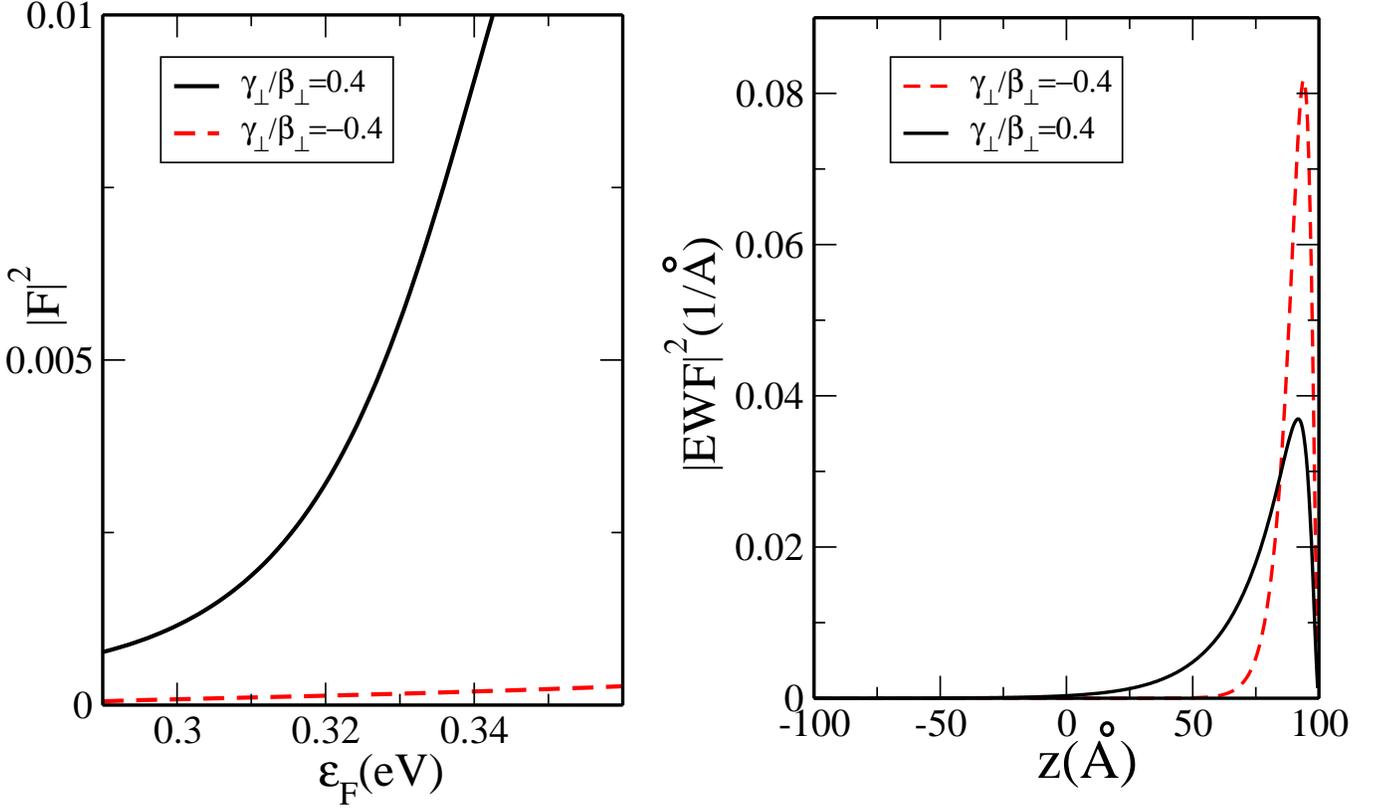}}
\caption{(Color online) (Left) Elastic (energy-conserving) surface-to-bulk transition matrix element calculated numerically as a function of  $\epsilon_F$. 
We take the quantum well state closest to the bulk band edge.
The solid (dashed) line corresponds to the case where the Dirac point is closer to the bulk conduction (valence) band.
(Right) Surface state envelope wave functions (EWF)  for a fixed energy in the conduction band,  when the Dirac point is close to the conduction band (solid black line) or the valence band (red dashed line).} 
 \label{fig:apn2}
\end{figure}

\subsection{F.~Temperature-dependence of the surface-to-bulk scattering rate induced by bulk acoustic phonons}

The objective of this section is to derive the temperature-dependence of the inelastic (phonon-induced) SB scattering rate.
When the temperature exceeds the largest phonon frequency ($\omega_0$), the dependence of $\Gamma_{\rm SB}^{\rm ph}$ on $T$ is clearly linear regardless of the type of phonon.
At temperature $T<\omega_0$, the optical phonon contribution is exponentially suppressed and the scattering rate is mainly dominated by acoustic phonons. 
In order to find out the power law of $T$, we consider an electron in surface band $c$ with momentum $k_F$ at the Fermi level scattering onto the $n'$-th quantum well state in the conduction band.
The rate for this process is
\begin{equation}
\label{lowt}
\Gamma_{{\bf k}_\parallel c\to c\, n'}\propto \sum_{s=\pm}\int d {\bf q}\, |g^{\rm ph}_{\bf q}|^2 |F_{{\bf k}_\parallel c, {\bf k}_\parallel-{\bf q}_\parallel\, q_z c\, n'}|^2 [n(\omega_{\bf q})+f(\omega_{\bf q})]\delta(\xi_{{\bf k_{\parallel}-\bf q_{\parallel}} 1n'}+ s\omega_{\bf q})
\end{equation}
where ${\bf q}=({\bf q}_\parallel, q_z)$, $\omega_{\bf q}=c_s\sqrt{q_{\parallel}^2+q_z^2}$, and 
\begin{equation}
|F_{{\bf k}_\parallel c, {\bf k}' c\, n'}|^2\equiv |F_{{\bf k}_\parallel c, {\bf k}' 1\, n'}|^2+|F_{{\bf k}_\parallel c, {\bf k}' 2\, n'}|^2 = \left|\int dz~  (\Psi_{{\bf k_{\parallel}}c}^S)^* e^{i k'_z z} \Psi_{{\bf k'_{\parallel}}1n'}\right|^2+ \left|\int dz~  (\Psi_{{\bf k_{\parallel}}c}^S)^* e^{i k'_z z} \Psi_{{\bf k'_{\parallel}}2n'}\right|^2.
\end{equation}
The phonon modes with $\omega_{\bf q}\gg T$ are exponentially suppressed and do not contribute to Eq.~(\ref{lowt}).
The main contribution comes from modes with $\omega_{\bf q}<<T$, for which $n(\omega_{\bf q})+f(\omega_{\bf q})\simeq T/\omega_{\bf q}$.
In order to estimate this contribution, we approximate Eq.~(\ref{lowt}) as
\begin{equation}
\label{lowt2}
\Gamma_{{\bf k}_\parallel c\to c n'}\propto T \int_{0}^{T/c_s} dq_\parallel q_\parallel \int_0^{2\pi} d\theta \int_{-T/c_s}^{T/c_s} d q_z |F_{{\bf k}_\parallel c, {\bf k}_\parallel-{\bf q}_\parallel\, q_z c\, n'}|^2  \delta(\xi_{{\bf k_{\parallel}-\bf q_{\parallel}} 1n'}\pm\omega_{\bf q}),
\end{equation}
where $\theta$ is the relative angle between ${\bf q}_\parallel$ and ${\bf k}_\parallel$, and we have used that $|g^{\rm ph}|^2/\omega_{\bf q}$ is independent of ${\bf q}$ for acoustic phonons.
Under the condition $\epsilon_F\gg T\gg\omega_{\bf q}$, it is appropriate to neglect $\omega_{\bf q}$ inside the delta function.
Then, Eq.~(\ref{lowt2}) transforms into
\begin{eqnarray}
\label{low3}
\Gamma_{{\bf k}_\parallel c\to c n'}
&\propto& T\int_{0}^{T/c_s}   dq_{\parallel} q_{\parallel} \int_0^{2\pi} d\theta    \int_{-T/c_s}^{T/c_s} dq_z |F_{{\bf k}_\parallel c, {\bf k}_\parallel-{\bf q}_\parallel\, q_z c\, n'}|^2  \delta(\xi_{{\bf k_{\parallel}-\bf q_{\parallel}} 1n'})\nonumber\\
&=& T\int_{0}^{T/c_s}   dq_{\parallel} q_{\parallel} \int_{-T/c_s}^{T/c_s} dq_z \int_{-1}^{1}  \frac{d\nu}{\sqrt{1-\nu^2}} |F_{{\bf k}_\parallel c, {\bf k}_\parallel-{\bf q}_\parallel\, q_z c\, n'}|^2  \sum_{i}\frac{\delta(\nu-\nu_i)}{| f'(\nu)|_{\nu=\nu_i}}.
\end{eqnarray}
In the second line of Eq.~(\ref{low3}), we have made a change in variable from $\theta$ to $\nu=\cos\theta$ and have rewritten the delta function in such a way that $\xi_{{\bf k_{\parallel}-\bf q_{\parallel}} 1n'}=0$ when $\nu=\nu_i$.
Note that $\nu_i$'s are functions of $k_{\parallel}$ and $q_{\parallel}$.
Also, $f$ is basically $\xi_{{\bf k_{\parallel}-\bf q_{\parallel}} 1n'}$ in terms of $\nu$, while $'$ denotes a derivative with respect to $\nu$.

When $q_\parallel=q_z=0$, $F_{{\bf k}_\parallel c, {\bf k}_\parallel-{\bf q}_\parallel\, q_z c\, n'}$ vanishes due to orthogonality between bulk and surface states.
At low temperatures, $q_\parallel$ is forced to be small compared to the Fermi wave vector and hence one may expand $F$ in powers of $q_\parallel$.
Likewise, when $T\ll c_s \lambda_s$, where $\lambda_s$ is the inverse penetration depth of the surface states into the bulk (cf. Sec. C), a small $q_z$ expansion of $F$ is well-justified. 
Then, we may write
\begin{equation}
\label{eq:exp}
|F_{{\bf k}_\parallel c, {\bf k}_\parallel-{\bf q}_\parallel\, q_z c\, n'}|^2\simeq a_{{\bf k}_\parallel} q_z^2+ b_{{\bf k}_\parallel} q_\parallel^2 \cos\theta^2.
\end{equation}
For $\lambda_s\simeq 1 {\rm nm}^{-1}$ and $c_s\simeq 2 {\rm km/s}$, the $q_z$ expansion is valid for $T\lesssim 15 {\rm K}$.
Incidentally, this temperature scale is similar to the Bloch-Gr\"uneisen temperature $T_{\rm BG}\equiv 2 c_s k_F$, for $k_F$ values of interest (cf. Fig. 1 in the main text).
When $T\gg T_{\rm BG}$, the small $q_z$ expansion is dubious but it can be shown that $\Gamma\propto T$.
In other words, the linear $T$ behavior starts to emerge even below the Debye temperature (note that the typical Debye temperature far exceeds $T_{\rm BG}$ in this system).
In the regime $T\ll T_{\rm BG}$, an explicit calculation shows that
\begin{equation}
\label{lowt4}
\Gamma_{{\bf k}_\parallel c\to c n'}\propto T\int_{q_-}^{T/c_s}   dq_{\parallel} q_{\parallel} \int_{-T/c_s}^{T/c_s} dq_z \frac{1}{q_\parallel}(a_{{\bf k}_\parallel} q_z^2 +b_{{\bf k}_\parallel}),
\end{equation}
where we have used  $T\gg c_s q_-$ and $\nu_i\propto q_\parallel^{-1}$.
Here, $q_-$ is the distance (in momentum space) between the surface state and the nearest bulk state at the Fermi level, while $k_F$ is the Fermi momentum.
The emergence of $q_-$ in the integral over $q_\parallel$ results from the fact that $\nu_i$ must lie between $-1$ and $1$. 
Clearly, the $a_{{\bf k}_\parallel} q_z^2$ term in Eq.~(\ref{lowt4}) gives a higher-power (i.e. subleading) contribution.
Then, in this temperature regime we obtain $\Gamma\propto T^3$.
If $T<c_s q_-$, we immediately obtain $\Gamma=0$.

Similarly, one can show that the intrasurface scattering rate induced by bulk acoustic phonons varies as $T^3$ below the Bloch-Gr\"uneisen temperature.
In order to see this, it is key to notice that $|F|^2$ for intrasurface scattering is {\em nonzero} at $q_z=q_\parallel=0$, because a surface state is not orthogonal to itself. 
The main difference between surface-to-bulk and intrasurface inelastic scattering is that $q_-=0$ for the latter. 
Hence, for intrasurface phonon scattering, the $T^3$ behavior continues all the way to $T=0$.


\end{widetext}

\end{document}